\documentclass[12pt,preprint]{aastex}
\newcommand{\beq}{\begin{equation}}
\newcommand{\eeq}{\end{equation}}
\def\stacksymbols #1#2#3#4{\def\theguybelow{#2}
        \def\verticalposition{\lower#3pt}
        \def\spacingwithinsymbol{\baselineskip0pt\lineskip#4pt}
        \mathrel{\mathpalette\intermediary#1}}
\def\intermediary #1#2{\verticalposition\vbox{\spacingwithinsymbol
        \everycr={}\tabskip0pt
        \halign{$\mathsurround0pt#1\hfil##\hfil$\crcr#2\crcr
                \theguybelow\crcr}}}

\catcode`\@=11
\def\gsim{\ifmmode{\mathrel{\mathpalette\@versim>}}
    \else{$\mathrel{\mathpalette\@versim>}$}\fi}
\def\lsim{\ifmmode{\mathrel{\mathpalette\@versim<}}
    \else{$\mathrel{\mathpalette\@versim<}$}\fi}
\def\@versim#1#2{\lower 2.9truept \vbox{\baselineskip 0pt \lineskip 
    0.5truept \ialign{$\m@th#1\hfil##\hfil$\crcr#2\crcr\sim\crcr}}}
\catcode`\@=12

\def\brem{bremsstrahlung$\;\,$}

\def\Msun{M_{\odot}}

\def\eps{\epsilon}
\def\epsz{\epsilon_0}

\def\epsw{\eps_{\rm w}}
\def\epswM{\epsw^{\rm M}}
\def\epsj{\eps_{\rm j}}

\def\lbh{L_{\rm BH}}
\def\lx{L_{\rm X}}

\def\ledd{L_{\rm Edd}}
\def\ldwin{L_{\rm dw}}

\def\lbhefUV{L_{\rm BH,UV}^{\rm eff}}
\def\lbhefopt{L_{\rm BH,opt}^{\rm eff}}

\def\lir{L_{\rm IR}}

\def\mast{M_*}

\def\mgas{M_{\rm gas}}

\def\mbh{M_{\rm BH}}

\def\Medd{M_{\rm Edd}}

\def\mdot{\dot\mbh}

\def\re{R_{\rm e}}

\def\tx{T_{\rm X}}

\def\etaw{\eta_{\rm w}}
\def\etawM{\etaw^{\rm M}}

\def\vw{v_{\rm w}}

\def\DOmew{\Delta\Omega_{\rm w}}

\def\BZw{B0$^{\rm w}$}
\def\BZzd{B0$_{02}$}
\def\BZzdw{B0$_{02}^{\rm w}$}

\def\BIw{B1$^{\rm w}$}
\def\BIzd{B1$_{02}$}
\def\BIzdw{B1$_{02}^{\rm w}$}

\def\BIIw{B2$^{\rm w}$}
\def\BIIzd{B2$_{02}$}
\def\BIIzdw{B2$_{02}^{\rm w}$}

\def\BIIIw{B3$^{\rm w}$}
\def\BIIIzd{B3$_{02}$}
\def\BIIIzdw{B3$_{02}^{\rm w}$}
\def\t15{t_{15}}

\shorttitle{Combined AGN feedback in ellipticals}
\shortauthors{Ciotti et al.}

\begin{document}
\slugcomment{Accepted version - May 13, 2010}

\title{Feedback from central black holes in elliptical galaxies.\\ III:
       models with both radiative and mechanical feedback}

\author{Luca Ciotti\altaffilmark{1},   
Jeremiah P. Ostriker\altaffilmark{2,3} \&
Daniel Proga\altaffilmark{4}}
\affil{$^1$Department of Astronomy, University of Bologna,
via Ranzani 1, I-40127, Bologna, Italy} 
\affil{$^2$Princeton University Observatory, Princeton, NJ, USA}
\affil{$^3$IoA, Cambridge, UK}
\affil{$^4$Department of Physics and Astronomy, University of Nevada,
Las Vegas, NV, USA}

\begin{abstract} 

  We find, from high-resolution hydro simulations, that winds from AGN
  effectively heat the inner parts ($\approx 100$ pc) of elliptical
  galaxies, reducing infall to the central black hole; and radiative
  (photoionization and X-ray) heating reduces cooling flows at the kpc
  scale. Including both types of feedback with (peak) efficiencies of
  $3\,10^{-4}\lsim\,\epsw\,\lsim 10^{-3}$ and of $\eps_{\rm EM}\simeq
  10^{-1.3}$ respectively, produces systems having duty-cycles,
  central black hole masses, X-ray luminosities, optical light
  profiles, and E+A spectra in accord with the broad suite of modern
  observations of massive elliptical systems.  Our main conclusion is
  that mechanical feedback (including all three of energy, momentum
  and mass) is necessary but the efficiency, based on several
  independent arguments must be a factor of 10 lower than is commonly
  assumed. Bursts are frequent at $z>1$ and decline in frequency
  towards the present epoch as energy and metal rich gas are expelled
  from the galaxies into the surrounding medium.  For a representative
  galaxy of final stellar mass $\simeq 3\,10^{11}\Msun$, roughly
  $3\,10^{10}\Msun$ of recycled gas has been added to the ISM since
  $z\simeq 2$ and, of that, roughly 63\% has been expelled from the
  galaxy, 19\% has been converted into new metal rich stars in the
  central few hundred parsecs, and 2\% has been added to the central
  supermassive black hole, with the remaining 16\% in the form hot
  X-ray emitting ISM.  The bursts occupy a total time of $\simeq 170$
  Myr, which is roughly 1.4\% of the available time. Of this time, the
  central SMBH would be seen as an UV or optical source for $\simeq
  45$\% and $\simeq 71$\% of the time, respectively. Restricting to
  the last 8.5 Gyr, the burst occupy $\simeq 44$ Myr, corresponding to
  a fiducial duty-cycle of $\simeq 5\,10^{-3}$.

\end{abstract}


\section{Introduction}

In a previous paper (Ciotti, Ostriker \& Proga 2009, hereafter Paper
I), we described in detail the physical processes that are included in
our current hydrodynamical modelling of the co-evolution of a massive
elliptical galaxy that contains a central supermassive black hole
(hereafter SMBH). Important elements include gas shed by evolving
stars, cooling flow driven infall to the central regions of this gas
and the associated star bursts, accompanied by accretion onto the
central SMBH and followed by nuclear and galactic winds driven from
the galaxy. Feedback in both radiative and mechanical forms is taken
into account, with the sources being the central SMBH, SNII and UV
from the newly formed stars, SNIa from older populations, and
thermalization of stellar mass losses.

To our surprise in Paper I and in the companion Paper II (Shin,
Ostriker \& Ciotti 2010a) we found that despite the richness of the
modeling we could not adequately represent the co-evolution of
elliptical galaxies and their central SMBHs, using the conventional
physics and {\it either} purely radiative or purely mechanical
feedback from the central SMBHs.  In retrospective, this is not
surprising because both processes operate in Nature, so that
presumably both are required to produce outcomes in agreement with
observations.

We can summarize the main results of our previous work as follows:

1) SNIa are energetically quite important and will drive winds from
elliptical galaxies (e.g, Ciotti et al.~1991; Ciotti \& Ostriker 2007,
hereafter CO07) but are only effective on the kpc scale, where the gas
densities are low. They cannot prevent cooling flows and massive
accumulations of gas into the inner regions of medium to massive
ellipticals when their present-day rate is in accordance with the most
recent observational estimates and the time evolution follows the
current theoretical indications (e.g., see Pellegrini \& Ciotti 1998).

2) Radiative feedback from central SMBHs (primarily the X-ray
component) and the young star generated feedback consequent to central
star bursts (e.g. Thompson, Quataert \& Murray 2005) can balance and
consume the cooling flow gas at the $10^2-10^3$ pc scale, but they
will not sufficiently limit the growth of the central SMBHs.  These
processes - radiation feedback and energy input from stellar evolution
- regulate the starburst phenomenon (Ciotti \& Ostriker 2001, CO07).

3) Mechanical feedback from the central SMBH on the $10^1-10^2$ pc
scale, mediated by a nuclear jet and the Broad Line Region winds
(e.g., Binney \& Tabor 1995; Begelman \& Nath 2005; Begelman \&
Ruszkowski 2005; Di Matteo, Springel \& Hernquist 2005), is efficient
in limiting the growth of the SMBH, but, absent the processes noted in
Point 2 above, would leave elliptical galaxies with more central star
formation (fed either by cooling flows or mergers) than is observed
(Papers I and II).

Thus, we concluded that all three sets of processes 1, 2, and 3 acting
on three different radial scales are in fact required (and, of course,
as noted all three -- SNIa, AGN radiation, Broad Line Region winds --
are observed phenomena) to match what we know of the properties of
elliptical galaxies. Note that characteristic time-scales, relevant
from the observational point of view, are associated in a natural way
to the different processes mentioned above. In practice, while SNIa,
stellar mass losses and gas cooling in the central regions of the
galaxy drive the global evolution of the galaxy gas mass budget on
temporal scales of several Gyrs (determined by the stellar evolution
clock), AGN feedback acts on shorter time scales, as each major
feedback event usually spans $10^7-10^8$ yrs.  The simulations also
revealed that each major central outburst actually consist of several
feedback events, on time scales of $\approx 1$ Myr or less, i.e., the
sound crossing time of the central kpc-scale region of the
galaxy. Finally, the last characteristic time scale ($\approx 0.5-1$
Gyrs) is that of dissipation of feedback effects on galactic scales,
setted by the sound crossing time over the galaxy body.

The purpose of this paper is to refine those
conclusions, to show which {\it combined models} (i.e., in which both
radiative and mechanical feedback effects are allowed) best fit
observations, and finally to propose observational tests of the
overall picture. We recall that all the presented simulations
represent the evolution of an intermediate luminosity, isolated
elliptical galaxy (i.e., no external pressure is imposed at the numerical grid
outer boundary), while the case of a galaxy in a cluster will be the
focus of future works. We assume, following long standing
observational data and recent simulations, that the elliptical galaxy
in question was made earlier and is close to its current state when
the calculation begins at $z\simeq 2$ (e.g. Renzini 2006, Naab et al.~2007).

\begin{figure}
\includegraphics[angle=0,scale=0.74]{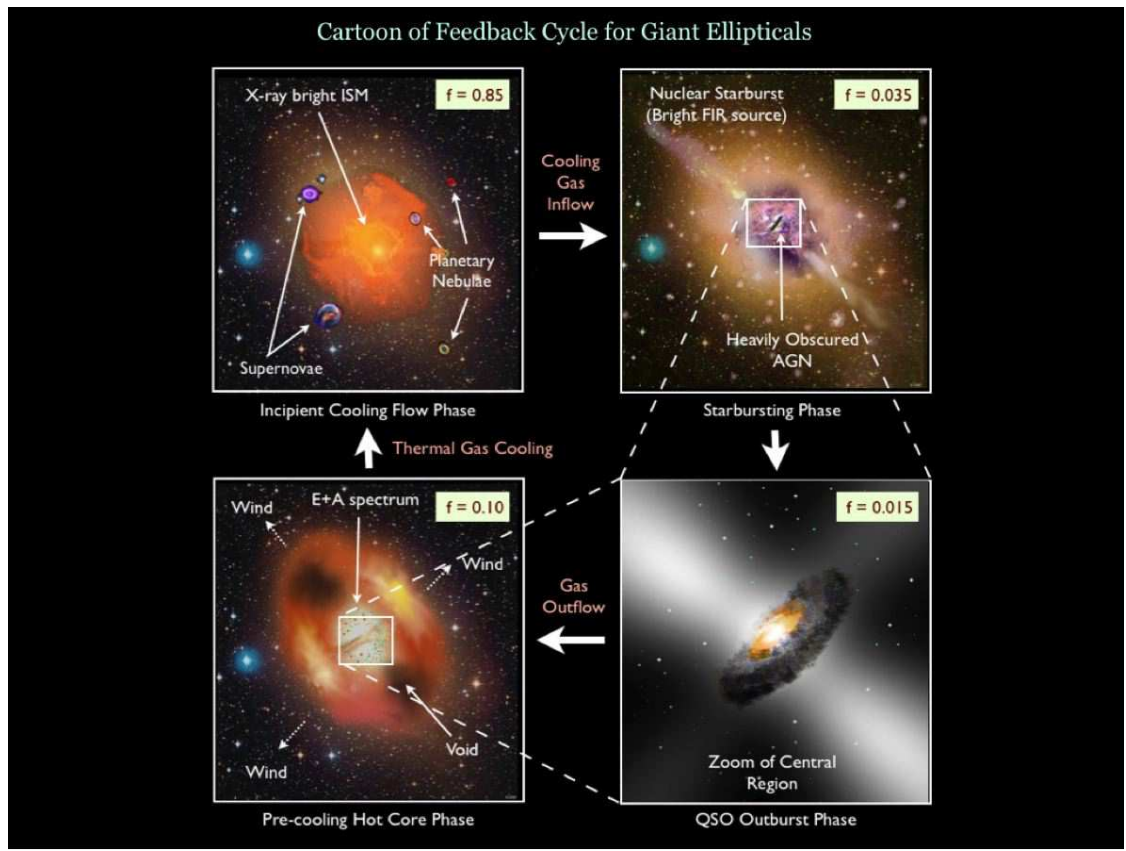}
\caption{This diagram shows the four main phases of the feedback cycle
  in the life of a galaxy. Secondary gas from stellar evolution leads
  to a cooling flow thermal instability that feeds a central SMBH, the
  outbursts from which leads to an expanding hot bubble which
  terminates the inflow.  The cycle may be repeated several times, and
  in each box we give the characteristic duty-cycle $f_{\rm duty}$ associated
  with each phase in a standard simulation.}
\label{fig:cartoon}
\end{figure}

The overall situation is inherently complex, and so, before diving
into the details of our new computations, it may be useful to present
a very rough, cartoon-level picture of the results. In
Figure~\ref{fig:cartoon} the arrows show the direction of time as the
galaxy/SMBH passes through four (of the many) phases of evolution,
and, since we are describing a cyclic phenomenon, we can start at any
point. A very rough estimate of the fraction of a cycle spent in any
given phase is shown in the sub-boxes at the upper right of each box
as a ``duty cycle'', $f_{\rm duty}$.

1) We arbitrarily begin at the upper left hand corner of the figure in
the more or less quiescent phase that occupies most of the time for
normal elliptical galaxies. Planetary nebulae and other sources of
secondary gas, processed through stellar evolution, are added to the
ambient gas everywhere in the galaxy at a rate proportional to the
stellar density and with an energy due to the stellar motions which
gurantees that, when the gas is thermalized, it will be approximately
at the local ``virial'' temperature without need for extra energy
input or output. Supernovae, primarily of type Ia are also distributed
like the stars and will tend to drive a mild wind from the outer parts
of the galaxy, with the inner parts being quite luminous in thermal
X-rays. This is a ``normal'' giant elliptical galaxy.

2) But, the gas in the dense inner part of the galaxy is radiating far
more energy than can be replaced by SNIa, stellar outbursts, cosmic
rays, conduction or any other energy source and thus a ``cooling
catastrophe'' occurs with a collapsing cold shell forming at $\approx
1$ kpc from the center. As this falls towards the center, a starburst
occurs of the type described by Thompson et al. (2005), and the galaxy
seen as an ULIRG.  A radio jet may be emitted, but the AGN flare up is
at first heavily obscured and the central source will only be seen in
hard X-rays.

3) Gradually, the gas is consumed, as it is transformed to new stars,
and some of it is driven out in a strong wind by the combined effects
of feedback from the starburst and the central SMBH, which is now
exposed as an optical and then UV ``quasar'', complete with Broad Line
Region (hereafter BLR) wind, optically thick disc of gas, and young
stars.

4) As gas is used up or blown away, a hot cavity is formed at the
center of the system and, since a shock has propagated through that
volume, it is essentially like a giant supernova remnant and one
expects there to be particle acceleration and non-thermal radiation
from the central region. This phase has been studied in detail in
Jiang et al. (2010). Then, gradually this hot bubble cools and
collapses and one returns to the normal elliptical phase at stage 1.

One is reminded of the Shakespearean seven ages of man in ``As You Like
It''. To paraphrase: ``And one galaxy, in its time, plays many parts'' -
the thermal X-ray source, the starburst, the quasar, the A+E system,
etc.

The paper is organized as follows. In Section 2, we briefly summarize
the main properties of the models adopted for the simulations,
referring to Paper I for all details of galaxy model construction and
input physics.  In Section 3, we present the results obtained when
adopting a fixed mechanical efficiency of the nuclear radiatively
driven wind. In Section 4, we describe in detail how the different
problems encountered in the models with fixed mechanical efficiency
are solved in the preferred class of simulations, in which the
mechanical feedback depends on the instantaneous accretion luminosity;
some relevant observational properties of these last types of model
are also described.  In Section 5, we also present preliminary results
obtained in a more advanced modelling of mechanical feedback.
Finally, in Section 6, we discuss the main results obtained,
observational tests and future developments.

\section{The models}

A full description of the galaxy models and the input physics adopted
for the simulations is given in Paper I and in Appendix of Paper II; a
comparative summary of the present and past treatments is given in
Table 1 of Paper I, while here we just recall the main properties of
the specific galaxy models used in the simulations.

For ease of comparison with the results of Papers I and II, we study
the main properties of a representative model characterized by an
initial stellar mass $\mast= 2.9\times 10^{11}\Msun$, a Fundamental
Plane effective radius $\re=6.9$ kpc, and a central aperture velocity
dispersion $\sigma_{\rm a}=260$ km s$^{-1}$, immersed in a dark matter
halo so that the total mass density distribution is proportional to
$r^{-2}$, with an identical amount of stellar and dark matter within
the spatial half-light radius.  The stellar distribution is modeled by
using a Jaffe (1983) profile\footnote{We correct a typo in Paper I.
  Just before equation (2) the correct expression is $\re=0.7447r_*$
  for the Jaffe profile. All the related numbers in the paper and in
  the code are correct.}, and all the dynamical properties needed in
the simulations are given in Ciotti, Morganti \& de Zeeuw (2009).  The
initial mass of the central SMBH is assumed to follow the present day
Magorrian relation ($\mbh\simeq 10^{-3}\mast$, see Magorrian et
al.~1998, see also Yu \& Tremaine 2000), as it is believed that the
bulk of the SMBH mass is assembled during the process of galaxy
formation (e.g., Haiman, Ciotti \& Ostriker 2004; Sazonov et
al.~2005), a process which is not addressed with the present
simulations. The initial conditions in this study are represented by a
very low density gas at the local thermalization temperature.  The
establishment of such high-temperature gas phase at early cosmological
times is believed to be due to a ``phase-transition'' when, as a
consequence of star formation, the gas-to-stars mass ratio was of the
order of 10\% and the combined effect of shocks, SN explosions and
AGN feedback became effective in heating the gas and driving galactic
winds. Several theoretical arguments and much empirical evidence, such
as galaxy evolutionary models and the metal content of the ICM support
this scenario (e.g., Renzini et al.~1993; Ostriker \& Ciotti 2005; Di
Matteo et al.~2005; Springel, Di Matteo, \& Hernquist 2005; Johansson,
Naab \& Burkert 2008).  For the reasons above, in the simulation here
presented (as well as in all others simulations not shown), we assume
that the age of the galaxy stellar component at the beginning of the
simulation is 2 Gyr old, and the simulations usually span ~12 Gyr, so
that the cosmic time at the end of the simulations is ~14 Gyr.

We set outflow boundary conditions at the galaxy outskirts ($\sim 250$
kpc), so that the simulations represent an isolated elliptical galaxy,
without the confining effect of the Intra Cluster Medium (ICM).  A
central cluster galaxy would have more difficulty generating winds and
would suffer from bursts of cluster gas inflow.  We adopted this
procedure to adhere to the standard approach followed in
``cooling-flow'' simulations, and to better evaluate the impact of the
combined feedback by comparison with the previously explored cases,
while in future explorations we will address in a more consistent way
the problem of the external pressure effects, of the galaxy structural
and dynamical modifications due to star formation and mass
redistribution over a Hubble time, the evolution of galaxies with
different initial mass and central velocity dispersion, and the
compatibility of the obtained galaxies with the present-day scaling
laws of elliptical galaxies (Ciotti 2009a).  Separately (Shin, Ostriker
\& Ciotti 2010b) we also illustrate the variations in evolutionary
paths caused by consideration of different initial mass galaxies and
differing in outer boundary conditions (in particular the stripping
process corresponding to different environments). We also stress that
the models here discussed are just a representative sample out of
several tens of runs that have been made, characterized by different
choices of the input physics parameters (often outside the currently
accepted ranges).

As in Paper I, the bolometric SMBH luminosity $\lbh$ produced by
accretion is related to the instantaneous accretion rate $\mdot$
(calculated according to equations [5]-[26] of Paper I) by a
luminosity-dependent electromagnetic (EM) efficiency $\eps_{\rm EM}$
as
\beq
\lbh =\eps_{\rm EM}\,\mdot\,c^2,\quad 
\eps_{\rm EM} =\epsz{A\dot m\over 1+A\dot m},\quad 
\dot m\equiv {\mdot\over\dot\Medd},
\eeq
where $\dot\Medd = \ledd/(\epsz c^2)$ is the Eddington mass accretion
rate.  $A$ is a free parameter so that the ``ADAF-like'' efficiency
scales as $\eps_{\rm EM}\sim \epsz A\dot m$ for $\dot m \ll A^{-1}$.
In our simulations we fix $A=100$ (see, e.g. Narayan \& Yi 1994; see
also Ciotti \& Ostriker 2001, where a very preliminary investigation
of ADAF effects on radiative feedback was carried out. Note that this
transition value seems also confirmed by recent observations, as
reported in Constantin et al.~2009).  We finally adopt as maximum
value for the EM efficiency $\epsz=0.1$ or 0.2 (e.g., see Noble,
Krolik \& Hawley 2009). We stress that in the treatment of radiation
feedback we consider, in addition to radiation pressure (whose
importance is well recognized, e.g, see Nayakshin \& Power 2010, King
2010, and references therein), calculated by solving the trasport
equation for the SMBH accretion luminosity and the starburst
luminosity (CO07 and Paper I), also heating/cooling (i.e., energy)
feedback (Sazonov, Ostriker \& Sunyaev 2004; Sazonov et al.~2005).

We now briefly mention the main aspects of the mechanical
feedback treatment that in the present {\it combined} models is added
to radiative feedback. In Paper I we introduced a nuclear wind (with
fixed opening angle in Type A models and with luminosity dependent
opening angle in Type B models), and a jet (with very small and fixed
opening angle), but the jet contribution was not included in the
models considered in Papers I and II.  Here we still neglect the jet
effects on the grounds that a thin relativistic jet will largely drill
through the central gas, depositing its energy at much larger radii,
and non-negligible effects are expected to be relevant only in the
low-luminosity, hot accretion phases characterizing late-time
evolution, perhaps by further reducing the nuclear
luminosity. However, as an additional piece of input physics we now
present some models in which the explicit time-dependent term in the
differential equation describing the mechanical feedback (Paper I,
equation [29]) is taken into account.

The common treatment of mechanical feedback (e.g., Di Matteo et
al.~2005, Johansson et al.~2008) 
is to estimate the mass inflowing to
the SMBH, multiply this by a coefficient $\epsw$ representing the
efficiency of conversion of mass to (kinetic) energy following the
prescription \footnote{Note that in our approach, we also consider the
  additional energy produced by stellar winds and SNII explosions in
  the stellar component of the circumnuclear disk, as apparent from
  equation (20) in Paper I.}
\begin{equation}
\ldwin=\epsw\dot\mbh c^2,
\end{equation}
and add that energy into the lower layers of the hydrodynamic
simulation in thermal form. Of course, in reality there is also radial
momentum added to the same zone via the outflowing wind that carries
the energy to the receiving layers. A recent paper (DeBuhr et
al.~2009) has focussed on the momentum in the radiation field, but we
know of no work that has allowed for the momentum associated with
mechanical energy emitted during AGN activity. This is curious since
in the parallel case of feedback associated with star formation it is
normally assumed that the momentum term dominates. But, as we noted in
Paper I, the consequence of the conservation equations can be that
much of the inflowing mass flows out again in the nuclear wind so that
the actual ratio of the SMBH accretion $\mdot$ to the rate of
inflowing mass ($\dot M_{\rm inflow}$, to be identified with $\dot
M^{\rm eff}_1$ in Paper I) is considerably less than unity. In fact,
from equations (14) and (22) in Paper I (see also Ostriker et al.~2010
for a full discussion) it is easy to show that the ratio is
\begin{equation}
{\mdot\over\dot M_{\rm inflow}} = {1\over 1+\beta},\quad\quad 
\beta={2\epsw c^2\over \vw^2},
\end{equation}
where $\vw\simeq 10^4$ km/s is the BLR wind velocity (for simplicity
here we neglect star formation in the circumnuclear accretion disk and
we assume stationarity).  For the low efficiency models which we will
later find most appropriate, (\BIIzd~and \BIIIzd~ in Section 4), the
ratio of the mass rates is roughly between 0.6 and 0.8. But for the
typical parameters used in the above quoted papers (i.e., $\epsw\sim
0.005$) this ratio can be $\simeq 0.1$ so that very little mass is
actually accreted on the SMBH if the calculation is done
consistently. As Paper I details, we allow for all three input
components (energy, momentum and mass). In a subsequent paper we will
show how omission of one or the other of these can greatly affect
results.

In the analysis of the numerical outputs, in addition to the
time-averaged quantities introduced in previous papers, i.e. the
accretion weighted electromagnetic and mechanical efficiencies
$<\eps_{\rm EM} >$ and $<\epsw >$ (Paper I, equation [33]), the
luminosity weighted nuclear wind opening angle $<\DOmew>$ (normalized
to the total solid angle, and restricted to the phases of the model
evolution when $\lbh>0.1\ledd$, Paper I, equation [34]), and finally
to the luminosity-weighted duty-cycle $f_{\rm duty}$ (Paper I,
equation [35]), we now also compute the {\it number of bursts} of each
model (each burst being counted when $\lbh$ becomes larger than
$\ledd/30$) and the total time spent at $\lbh \geq \ledd/30$
(bolometric). For selected models we also compute the number of bursts
and the total time spent at $\lbhefUV\geq 0.2\ledd /30$ (UV, after
absorption), and at $\lbhefopt\geq 0.1\ledd /30$ (optical, after
absorption). The two numerical coefficients take into account the
fraction of the bolometric luminosity used as boundary condition to
solve the radiative transfer equation in each of the two
bands. Therefore, we can now test how well the luminosity-weighted
$f_{\rm duty}$ used in our previous papers performs against the
estimates of duty-cycle obtained by direct number count (see Table 1).
And, more importantly, see which, if any models show a fraction of
time in the high state comparable to the fraction of black holes
observed to be in an AGN phase.

We finally recall that the stellar mass loss rate and the SNIa rate
associated with the initial stellar distribution are the main
ingredients driving evolution of the models. In the code the stellar
mass losses -- the source of {\it fuel} for the activity of the SMBH
-- follow the detailed prescriptions of the stellar evolution theory,
and we use exactly the same prescriptions as in CO07 (see Sects.~2.2
and 2.3 therein).  The radiative heating and cooling produced by the
accretion luminosity are numerically computed as in CO07 by using the
Sazonov et al.~(2005) formulae, which describe the net heating/cooling
rate per unit volume of a cosmic plasma in photoionization equilibrium
with a radiation field characterized by the average quasar Spectral
Energy Distribution derived by Sazonov et al.~(2004, see also Sazonov
et al.~2007, 2008), whose associated spectral temperature is
$\tx\simeq 2$ keV.  In particular, Compton heating and cooling, \brem
losses, line and continuum heating and cooling, are taken into
account. Also the star formation over the galaxy body, the radiation
pressure due to electron scattering, to phoionization, and finally to
UV, optical and infrared photons on dust, are treated as in CO07,
where the derivation and the numerical integration scheme of the
radiative transport equations is described in detail. All the relevant
information about the 1D numerical code and the hydrodynamical
equations can be found in Ciotti \& Ostriker (2001) and CO07.

\section{Combined models with fixed mechanical efficiency}

In order to better follow the description of the results of the new
simulations, here we summarize the principal phases of the model
evolution as produced by the numerical simulations. These main aspects
are almost independent of the specific modelization of the mechanical
feedback treatment, and so apply also to the models in Section 4.

Overall, we found that in combined models the main properties of model
evolution are preserved; in particular, episodic outbursts reaching
$0.1\ledd$ can be common.  After a first evolutionary phase in which a
galactic wind is sustained by the combined heating of SNIa and
thermalization of stellar velocity dispersion, the central ``cooling
catastrophe'' of the galaxy gaseous halo commences, with the formation
of a collapsing cold shell at $\sim 1$ kpc from the center.  In
absence of the feedback from the central SMBH a ``mini-inflow'' would
then be established, with the flow stagnation radius (i.e., the radius
at which the flow velocity is zero) of the order of a few hundred pc
to a few kpc: these decoupled flows are a specific feature of cuspy
galaxy models with moderate SNIa heating (Pellegrini \& Ciotti
1998). After the cooling catastrophe, the SMBH feedback affects the
subsequent evolution, and the number of bursts depends on the nature
and strenght of the feedback. Nuclear bursts are accompained by
significant episodes of star formation in the central regions of the
galaxy, and obscuration of the nuclear source can be considerable. As
already found in CO07, the major AGN outbursts are separated by
increasing intervals of time (set by the cooling time and by the
secular decrease of the mass return rate from the evolving stellar
population), and present a characteristic temporal substructure, whose
origin is due to the cooperating effect of direct and reflected shock
waves (from the inner rim of the spherical strongly perturbed
zone). In general, purely mechanical feedback tends to produce
``clean'' bursts, while radiative feedback leads to bursts consisting
of several sub-bursts. In all cases, these outflowing shocks are a
likely place to produce emission of synchrotron radiation and electron
and ionic cosmic rays (see Jiang, Ostriker \& Ciotti 2010, see also
Sijacki et al.~2008). Finally, at late epochs the galaxy models are
usually found in a state of stationary, very sub-Eddington hot
accretion, in an ADAF state.

We now illustrate the results obtained in case of combined models
starting with the description of the simplest combined models,
designed by ``A'' in Paper I.  These A models should be compared with
the corresponding purely mechanical MA models in Papers I (Table 2)
and II, in order to determine the additional effects of radiative
feedback; in fact MA models were meant to investigate the effects of
mechanical feedback, so that radiative effects were excluded from the
simulations.  Type A models adopt the commonly assumed prescription of
a fixed mechanical efficiency for the nuclear wind, i.e. in equation
(2)
\begin{equation}
\epsw=\epswM,
\end{equation}
where $\epswM$ is the value reported in Column 2 of Table 1. From
equation (2) it follows that the mechanical energy flowing out of the
central SMBH regions bears a fixed relation to the mass accreted by
the SMBH.  This is what is normally assumed in works on AGN feedback
(e.g., Di Matteo et al.~2005, Johansson et al.~2008).  Therefore, for
the A models the mass-accretion weighted nuclear wind efficiency
$<\epsw>$ (equation [33] in Paper I), coincides with $\epswM$, as
apparent from Column 3 of Table 1.  It is important to stress,
however, that in the simulations only a fraction of the mechanical
energy is transferred to the ISM, and at different radial scales, as
described by the physically based differential equation (29) derived
and discussed in detail in Paper I.  In all the presented models
A0-A3, the mass ejected with the nuclear wind (that would be observed
as a BLR wind), is a factor of two greater than the mass accreted on
the central SMBH, and the nuclear wind opening angle is maintained
fixed, i.e.
\begin{equation}
\etaw=2,\quad \DOmew=\pi ,
\end{equation}
(see equations [17], [18], and [31] in Paper I).  Finally, the maximum
radiative efficiency $\epsz$ in equation (1) adopted in the four A
models presented in Table 1 is fixed to 0.1: note how the implemented
ADAF-like treatment, leads to the reduced accretion-mass weighted
values for $<\eps_{\rm EM}>$, as reported in Column 5.

The model results are given in Table 1 in order of decreasing
mechanical efficiency $\epswM$ (Column 2), with the first model A0
adopting the conventionally chosen constant mechanical efficiency of
0.005; a direct comparison with the results of the correspondent
purely mechanical models MA0-MA3 can be done by inspection of Table 2
in Paper I.  As expected, reducing the wind mechanical efficiency
increases the SMBH accretion rate, and so the emitted luminosity
increases, as apparent from the values in Columns 5 and 6. Column 7,
where we list the total stellar mass of the new stars, confirms a
previous finding of our models: in practice, also in the combined
models {\it a larger number of AGN bursts (consequence of a reduction
  of the mechanical feedback efficiency), leads to a larger amount of
  star formation}.  In addition, the final mass of new stars is
(slightly) larger than in the purely mechanical MA models, due to the
fact - already described in Paper I - that single bursts in presence
of radiative feedback tend to be richer in the temporal substructure,
leading to a longer period of nuclear star formation
activity. Therefore we again confirm that, while it is believed that
in the process of galaxy formation the feedback from the central SMBH
may help to end the galaxy formation epoch, in the successive (and
unavoidable) evolution driven by the mass return from the evolving
stars, AGN feedback may actually lead to an increase of star
formation, in the inner few hundred parsecs of the galaxy. As
described in CO07, this phenomenon can lead to a build-up of a stellar
``cusp'' of new stars at the center of galaxies experiencing repeated
AGN burst, with structural properties in nice agreement with
observational findings.  We will return more extensively on this point
in Section 4.2.

As noted in Section 2 the model with the ``standard'' high efficiency,
$A0$, requires such a large fraction of the inflowing mass to be
returned in the ouflowing wind that very little accretes onto the
central SMBH. As a consequence its mass increases by only $\sim
10^7\Msun$ (Table 1, Column 6), and the observed $\mbh$-$\sigma$
relation could certainly not be maintained by this type of model. In
addition, the final thermal X-ray luminosity would be too low.

Remarkably, the time-integral of the gas lost by the galaxy at 10
$\re$ (Column 8), does not depend strongly on the assumed
value of mechanical feedback efficiency, as it is mainly driven by
SNIa heating. Actually, a {\it decrease} of the mechanical wind
efficiency produces a slightly larger degassing, a consequence of the
increased number of bursts. The combined effect of SMBH accretion,
galaxy degassing, and star formation is the explanation of the
non-monotonic behavior of the present-day amount of hot, X-ray
emitting gas in the galaxy (Column 9): for very high efficiencies
(model A0) the gas mass is very low, so that also the X-ray luminosity
is correspondingly low (Column 11). At very low mechanical
efficiencies (model A3), a significant amount of gas has been accreted
on the SMBH but, most importantly, the considerable star formation
induced by the repeated bursts deprives the galaxy of gas: for
intermediate values of mechanical efficiency (model A2) the amount of
hot gas is maximized, and also the X-ray luminosity reaches a
present-day value consistent with X-ray emitting galaxies. As already
stressed in the Introduction, we again recall that pressure from the
ICM will increase the luminosity values. In summary, matching the
models and the observed X-ray thermal luminosity is a sensitive
contraint. Both too little or too much feedback deprives the galaxy of
gas and reduces the X-ray output to values below what is observed.

In Column 10 we list the present-day SMBH luminosity, measured in
units of Eddington luminosity: the main reason for the decline at
decreasing mechanical efficiency is the substantial increase in the
final SMBH mass. In Column 13 we report the number of bursts of the
model.  We count as a burst every time the bolometric accretion
luminosity grows above the fiducial limit of $\ledd /30$, and in
Column 14 we report the total time (in Myrs) spent by the bolometric
accretion luminosity above the threshold of $\ledd/30$. As already
described, the number of bursts increases with decreasing mechanical
efficiency, and also the average duration of each burst, going from
$\approx 4$ Myr in model A1 up to $\approx 12$ Myr in model A3. These
results are in nice agreement with observational estimates 
(e.g., Kirkman \& Tyler 2008).
{\it It is important to stress that the obtained time-scales for the
  nuclear bursts are not imposed, but are self-determined by the
  evolutionary physics}: in addition, several sub-bursts are in
general grouped in a time extended single burst, following the physics
described in CO07 and in Paper I.

\begin{figure}
\includegraphics[angle=0,scale=0.8]{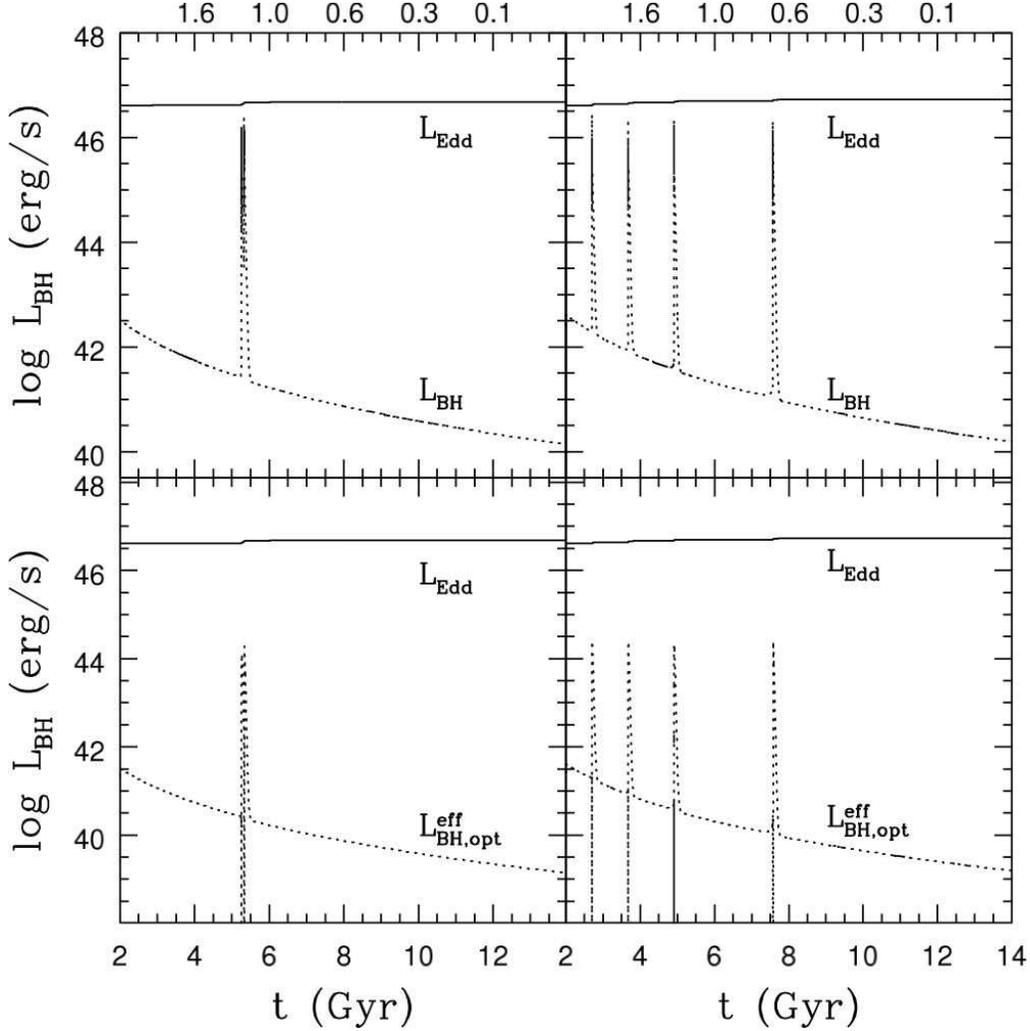}
\caption{Luminosity evolution of model A1 (left panels), a model with
  constant mechanical efficiency $\epsw=2.5\,10^{-4}$ and also
  radiative feedback. After one large burst the gas reservoir in the
  galaxy is so depleted that no further bursts occur. Same quantities
  for model A2 (right panels), in which $\epsw=10^{-4}$. Dotted lines
  are the bolometric accretion luminosity (top) and the optical SMBH
  luminosity corrected for absorption, i.e., as it would be observed
  from infinity (bottom). As in CO07 at the center we fixed
  $\lbhefopt(R_1)=0.1\lbh$.  The global properties of the models are
  given in Table 1.  In the top horizontal scale of this and of all
  the other figures the redshift corresponds to the bottom time
  scale.}
\label{fig:fla}
\end{figure}

For example, the 9 bursts counted in model A1 are just the temporal
substructure of the single burst appearing in Figure~\ref{fig:fla}
(left panels). In this model after one large burst most of the ISM in
the galaxy is expelled and there is little further activity. This is
similar to what Di Matteo et al. (2005) found, and has some attractive
features in showing the power possible from AGN feedback, but, as
noted above, very little mass will be accreted on the central SMBH in
this type of model if mass conservation is strictly enforced.  The
evolution of model A2, with a reduction of the mechanical efficiency
by a factor of 2.5 is also shown in Figure~\ref{fig:fla} (right
panels). The most evident and expected feature is the increase in the
number of bursts.

Are these models (A1 and A2) an adequate representation of reality?
Was the addition of radiative feedback able to cure some of the
problems affecting the purely mechanical MA models described in Papers
I and II?  We think that they are not. In fact, there are two salient
defects. First the final gas luminosity and the final gas mass are
both below those typically seen in elliptical galaxies (Pellegrini,
Ciotti \& Ostriker 2009; Diehl \& Statler 2007).  Second, as noted,
the central SMBH accretes very little mass. The reason for the much
lower accretion found here than in other works (e.g., Di Matteo et
al.~2005) is in part that the latter have assumed an accretion rate of
100 times the Bondi rate, whereas we have computed the accretion rate
self-consistently (see also a discussion of this issue in Kurosawa et
al.~2009, and Figure 1 therein); and in addition the outwards mass
flow has not been, in general, incorporated in the calculation. In any
case this high assumed wind efficiency seems to lead to results in
conflict with well established observations, so we have rescaled to
lower values the assumed wind efficiency in model A3, keeping the peak
radiative efficiency at 0.1, and the mechanical efficiency constant.
In conclusion, it does not appear that low efficiency models will be
satisfactory for any adopted value of the mechanical efficiency. The
reason is that the energy input is restricted to a region very near
the central SMBH and that models with sufficiently low efficiency to
grow reasonable size SMBHs have so little energy input into the bulk
of the galaxy that cooling flow induced starbursts leave the galaxy
with too much gas and too high rates of central star formation.  Also,
very low efficiency models would leave most ellipticals with an E+A
spectrum seen in the central regions due to recurrent starbursts
(e.g., see Wang et al.~2009, see also Paper II).

\section{Combined models with luminosity-dependent
 mechanical efficiency}

On the basis of the previous investigation, we are now in the position
to explore a family of models (type B models) in which the amount of
gas ejected by the nuclear wind, the mechanical wind efficiency, and
the wind opening angle depend on the Eddington normalized bolometric
accretion luminosity $l =\lbh/\ledd$ as
\begin{equation}
\etaw={3\etawM\over4}{l\over 1+0.25\,l},\quad
\epsw={3\epswM\over4}{l\over 1+0.25\,l},\quad
\DOmew=\pi\, {\rm min}(\sqrt{l^2 + a^2},1),
\end{equation}
where $a=2.5\,10^{-2}/\pi$ (Paper I, equations [18], [21], [31]; and
Figure 1 therein).  In order to have a velocity of nuclear winds with
observed values we must assume $\etawM\simeq 1800\epswM$, in
accordance with the relation
\begin{equation}
\vw\simeq\sqrt{2\epsw\over\etaw}\,c\approx 10^4\,{\rm km}\,{\rm s}^{-1}.
\end{equation}
These assumptions, while somewhat different from those normally made
by galaxy modellers, are closer to what is expected from studies of
central engines (e.g., Proga, Stone \& Kalman 2000; Proga \& Kalman
2004, Benson \& Babul 2009), and possibly with observational evidences
(Allen et al.~2006).  In particular, wind efficiency and ejected mass
fraction increase at increasing $l$ and saturate at $l\geq 2$ at the
values $\epswM$ and $\etawM$; the wind (solid) opening angle also
increases for increasing $l$, in the range $2.5\,10^{-2}$ to $\pi$.
As in the case of combined A models, also in combined B models we
consider both mechanical and radiative feedback, adopting the ADAF
prescription (1). However, we now explore two families of radiative
efficiencies: in the first (analogous to A models presented in the
previous Section) $\epsz=0.1$, while in the second (the B$_{02}$
models in Table 1) the peak EM efficiency is increased to $\epsz=0.2$.
Again, as for A models, the explicit time dependent term in the
differential equation describing the discharge of the nuclear wind
mass, momentum, and kinetic energy on the galaxy ISM (Paper I,
equation [29]), is not taken into account.  This further ingredient is
activated in the suite of B$^{\rm w}$ models, briefly described in the
following Section.

We start by considering the overall properties of the combined B
models with $\epsz=0.1$ (B0-B3 in Table 1), ordered for decreasing
$\epswM$. A few systematic behaviors can be easily spotted. For
example, as for A models, the mass accretion weighted mechanical
efficiency $<\epsw>$ (Table 1, Column 3) decrease moving from
model B0 to model B3, a direct consequence of the reduction in the
adopted value for the peak mechanical efficiency $\epswM$. Also, the
mass accretion weighted EM efficiency $<\epsilon_{\rm EM}>$ increases
at decreasing $\epswM$, due to the increasing number of burst events,
and so of high luminosity peaks (Column 13).  The total mass accreted
by the SMBH, $\Delta\mbh$, as expected increases from model B0 to
model B3. However, the variation of $\Delta\mbh$ for different values
of $\epswM$ is smaller than in the A family. In general, for similar
values of $\epswM$, the accreted mass in B models is larger than in A
models: this difference is due to the fact that in B models, between
two successive bursts, the mechanical feedback drops because the solid
opening angle decreases to values corresponding to a classical jet,
and the mechanical efficiency drops virtually to zero.  This is not a
minor effect, and it shows how a non-negligible amount of mass is also
accreted in the low-luminosity phases between bursts.  The amount of
star formation also increases at decreasing $\epswM$, as for the
family of A models (Table 1, Column 7). But, as for the total mass
accreted on the SMBH, also for the mass in new stars B models
accumulate significantly larger masses, with final $\Delta\mast\gsim
10^9\Msun$ in B0, up to $10^{10}\Msun$ in model B3.

The additional effect of radiative feedback on the models with a
luminosity dependent mechanical feedback can be determined by
comparison of the present results with those of purely mechanical MB
models listed in Table 2 of Paper I. Here we just notice how the final
SMBH masses in combined models are in better agreement with
observational values, where purely mechanical (and the purely
radiative RB models, also presented in Table 2 of Paper I) models tend
to produce too large SMBHs. This fact clearly points out the
importance of the co-operation of mechanical and radiative feedback in
self-regulating galaxy evolution. This point will be discussed on more
physical grounds in Section 4.2.  The effects of an increase of the
peak value of radiative efficiency in combined models (from 0.1 to
0.2) are shown by the B$_{02}$ family in Table 1. All the trends with
$\epswM$ are similar to models with $\epsz=0.1$, but the final SMBH
masses slightly reduced, as expected.

Overall, we can summarize the main results of this Section as follows.
It is apparent that combined B models behave better than combined A
models, in the sense that mechanical feedback is sufficiently strong
to add its effects to radiative feedback, but it is not so strong as
to prevent the recurrent cooling catastrophe events, so that the
galaxy is not in a permanent wind state, and contains sufficient gas
to produce substantial coronal X-ray luminosity.  In the case of very
low assumed efficiency for wind input, of course the global evolution
is similar to that of purely radiative models, while in the case of
high efficiency the evolution is significantly different.  Motivated
by the encouraging results of the combined B family, in the following
Section we focus our attention on two B models that account most
satisfactorily for several observational properties. We recall that
details of X-ray properties of these models (such as their surface
brightness in the soft and hard bands used in observational works,
luminosity weighted X-ray temperatures during the different
evolutionary phases) have been already presented elsewhere
(Pellegrini et al.~2009).

\subsection{Representative ``best'' models}

The two ``best'' models among those we have studied are
\BIIzd~($\epswM=10^{-3}$) and \BIIIzd~($\epswM=3\,10^{-4}$), with peak
mechanical efficiencies in agreement with observational values
(Schlesinger et al.~2009). These will be considered our fiducial
models bracketing the optimal values for the input parameters and for
output quantities. Each of the two has interesting properties that we
now discuss in some detail, beginning with a discussion of the
time-evolution of luminosities and mass accretion rates.

\begin{figure}
\includegraphics[angle=0,scale=0.8]{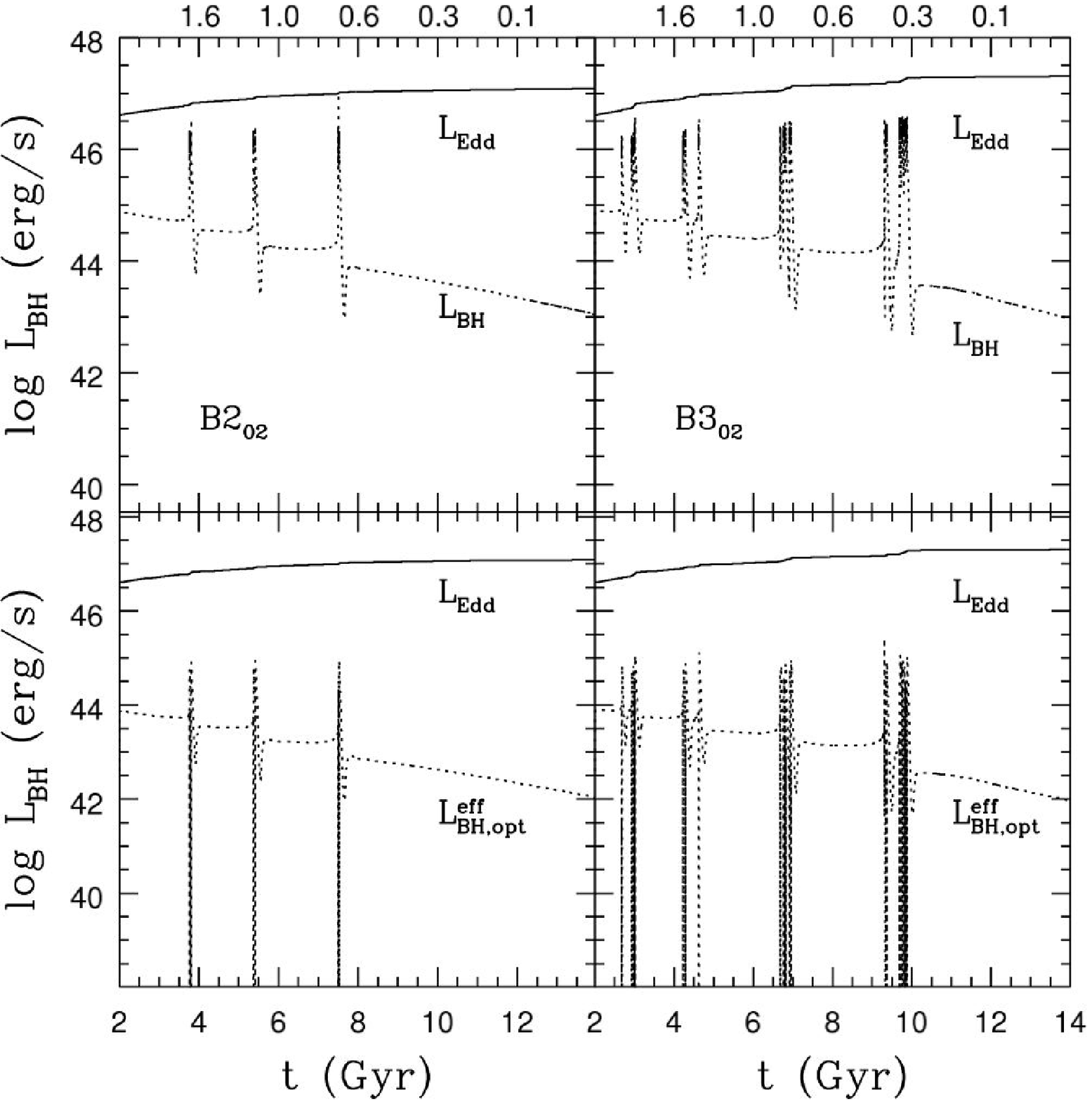}
\caption{Fiducial models showing evolution of SMBH luminosity.  Luminosity
  evolution of model \BIIzd~(left panels), a model with peak
  mechanical efficiency $\epswM=10^{-3}$ and also radiative feedback.
  Same quantities for model \BIIIzd~(right panels), in which
  $\epswM=3\,10^{-4}$. Dotted lines are the bolometric accretion
  luminosity (top) and the optical SMBH luminosity corrected for
  absorption, i.e., as it would be observed from infinity
  (bottom). The global properties of the models are given in Table 1.
  Note that at the current epoch ($z=0$) the SMBH bolometric
  luminosity $\lbh$ is four orders of magnitude below the Eddington
  limit $\ledd$.}
\label{fig:flb}
\end{figure}
\begin{figure}
\includegraphics[angle=0,scale=0.8]{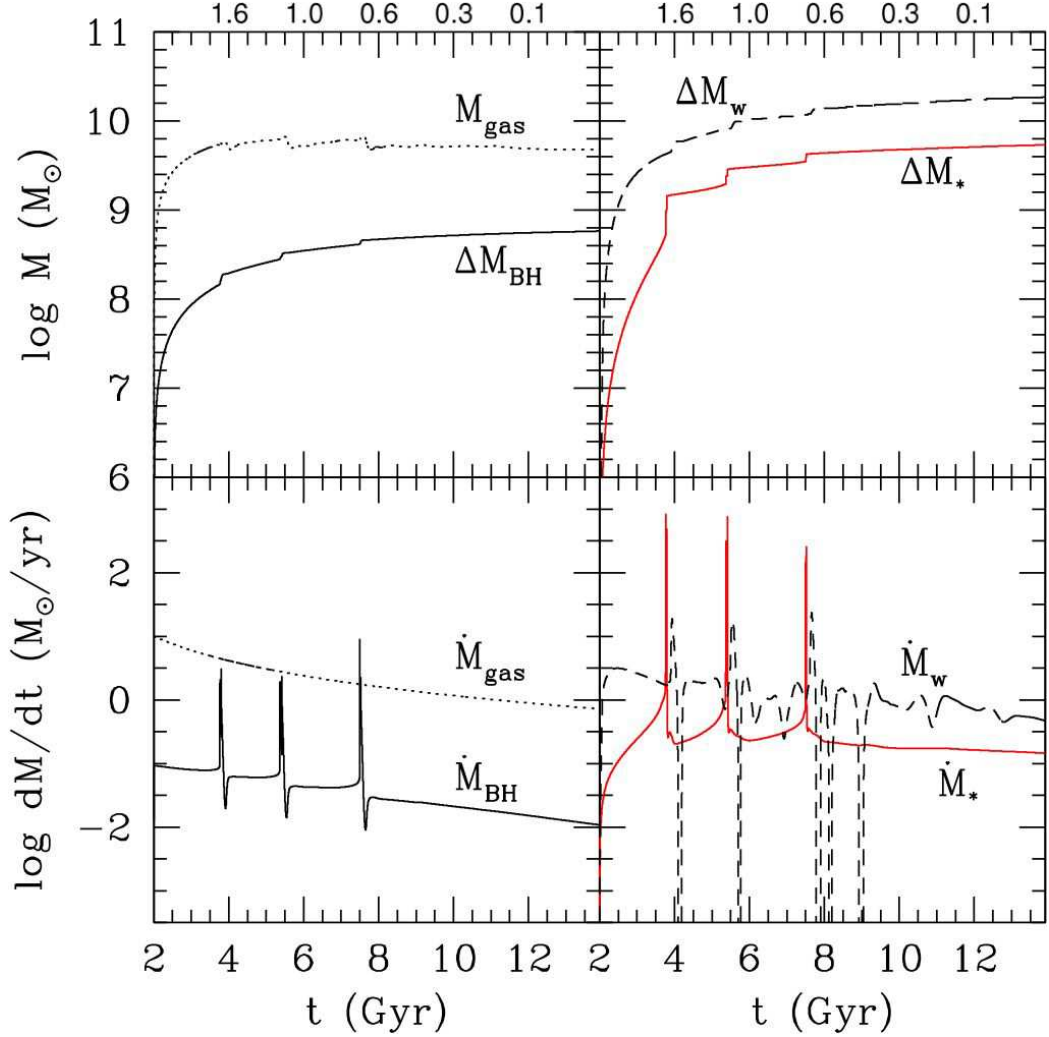}
\caption{Mass budget evolution of model \BIIzd. Top left panel: mass
  accreted on the central SMBH and ISM mass inside $10\re$.  Top
  right: total mass of gas ejected from the galaxy at $10\re$, and
  total mass of new stars in the galaxy.  In the bottom panels the
  corresponding rates are shown.  The bulk ($\simeq 63\%$) of the
  recycled gas is ejected in galactic
  winds, with $\simeq 19\%$ of it converted to centrally located
  stars, $\simeq 2\%$ of it accreted to the central SMBH, and finally
  $\simeq 16\%$ of it under the form of X-ray emitting hot ISM.  Note
  that in the pulsed activity of bursts the wind ejection rate from
  the galaxy can exceed 100 $\Msun$/yr, and also how peaks in the
  star formation anticipate the large degassing events.}
\label{fig:fmb202}
\end{figure}
\begin{figure}
\includegraphics[angle=0,scale=0.8]{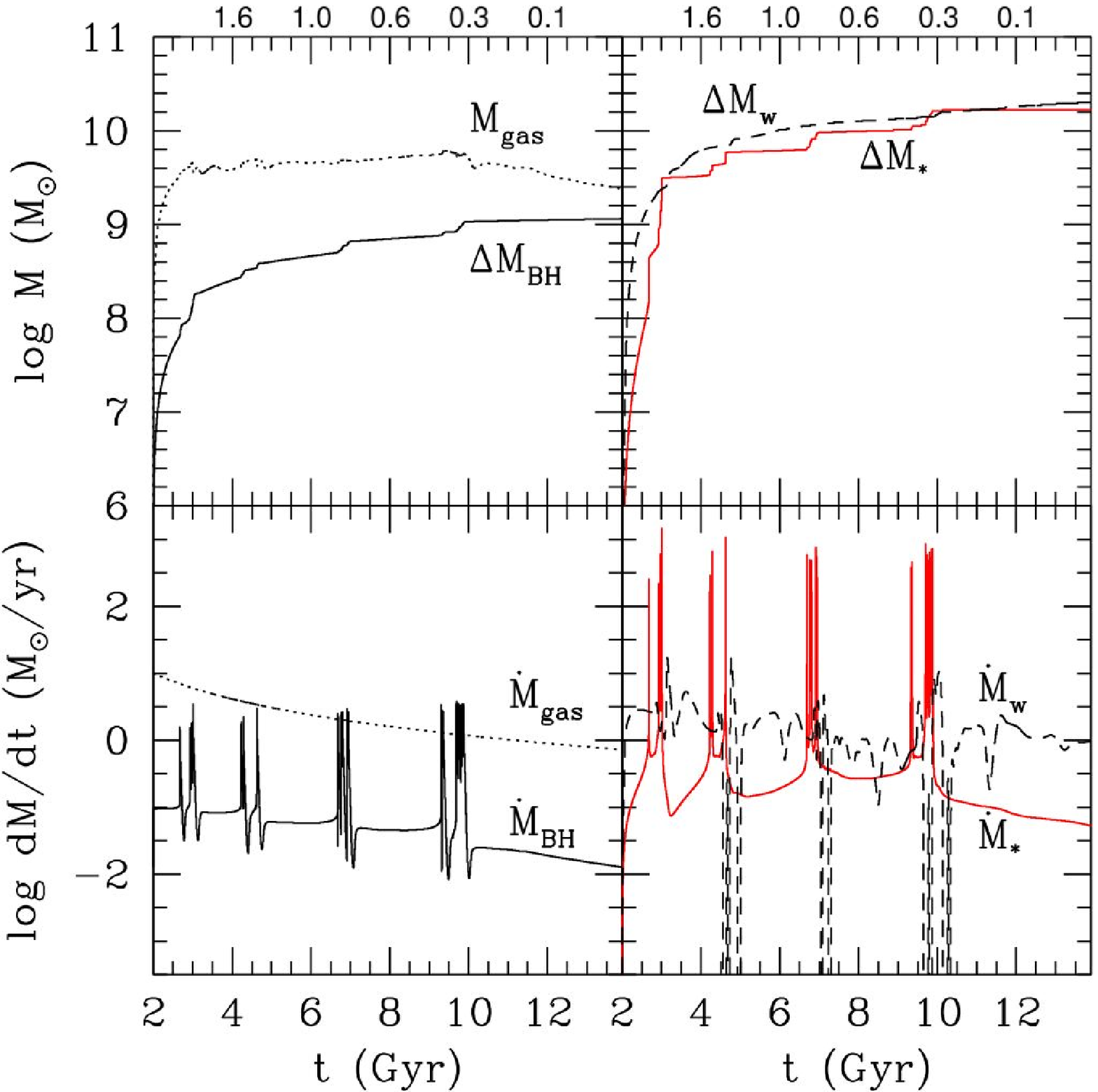}
\caption{Mass budget evolution of model \BIIIzd. Panels and lines as
  in Figure~\ref{fig:fmb202}.  Behavior is essentially similar but the
  weaker mechanical feedback permits bursts to last till later cosmic
  time.  The frequency and intensity of mass expulsion and star
  formation events is paradoxically greater for lower mechanical
  efficiency.}
\label{fig:fmb302}
\end{figure}

The luminosity evolution of the two models is shown in
Figure~\ref{fig:flb}. From comparison of the left and right panels, it
is apparent how a reduction in the peak value of mechanical efficiency
affects the time-evolution, number, and temporal structure of the
bursts. Since the mechanical efficiency in model \BIIzd~is three times
higher than in model \BIIIzd, the number of bursts in the former is 3,
while in the second 4; moreover, as common in radiative feedback
models (see also Paper I) each single burst is organized in several
sub-bursts. We note that the central optical/UV luminosities are far
below the Eddigton limit (five orders of magnitude!) at the current
epoch in these fiducial models in rough agreement with current
observations (e.g., see Pellegrini 2005, Ho 2009).  However, we find
that even at these accretion rates the mechanical and radiative
feedback is sufficiently strong to heat gas near the SMBH to the level
when the numerical self-consistently calculated ``Bondi-rate'' is very
low.

This luminosity evolution corresponds to the different mass accretion
(and ejection) histories, as plotted in Figures~\ref{fig:fmb202}
and~\ref{fig:fmb302}. In fact, the final SMBH mass and mass of the new
stars added to the galaxy is $\Delta\mbh\simeq 5.9\,10^8\Msun$ and
$\Delta\mast\simeq 5.5\,10^9\Msun$ in model \BIIzd, while for the
model \BIIIzd~the same quantities are $\simeq 1.1\,10^9\Msun$ and
$\simeq 1.7\,10^{10}\Msun$, respectively (see Table 1). The ejected
mass as a galactic wind is very similar in both cases, summing up to
$\simeq 2\,10^{10}\Msun$ of material. The star formation rate during
the periods of feedback dominated accretion oscillates from $0.1$ up
to several hundreds (with peaks near $10^3$) $\Msun$ yr$^{-1}$ (the
ULIRGs level), while it drops monotonically to $\lsim 10^{-1}$ $\Msun$
yr$^{-1}$ in the last 6 Gyrs of quiescent accretion in model
\BIIzd~(Figure \ref{fig:fmb202}, red line, bottom panel), and to $\sim
3\,10^{-2}$ $\Msun$ yr$^{-1}$ in the last 4 Gyrs in model
\BIIIzd~(Figure~\ref{fig:fmb302}), consistently with observations of
star formation signatures in optically quiescent early-type galaxies
(e.g., Salim \& Rich 2010).  It is also apparent how the star
formation episodes are on one side enhanced by SMBH feedback, but also
how they end abruptly after major SMBH outburts (e.g., see 
Schawinski et al.~2009, Nesvadba et al.~2010).

\begin{figure}
\includegraphics[angle=0,scale=0.8]{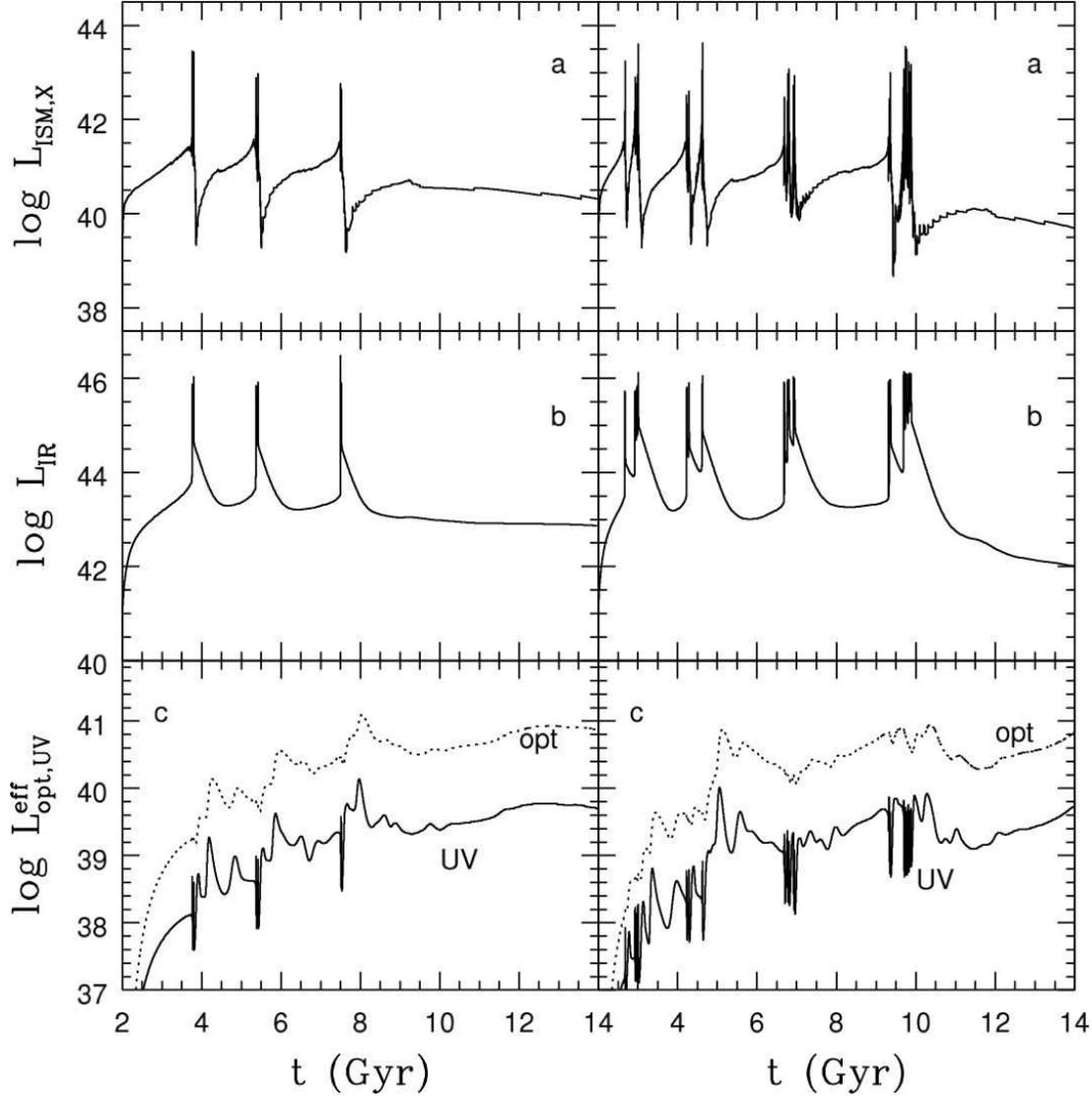}
\caption{Time evolution of the galactic X-ray coronal luminosity $\lx$
  (top), recycled infrared luminosity $\lir$ (middle), and the
  starburst UV and optical luminosities (bottom), corrected for
  absorption. Left panels refer to model \BIIzd~ and right panels to
  model \BIIIzd.  The infrared luminosity is due to the reprocessing
  of the radiation emitted by the new stars and by the SMBH and
  absorbed by the ISM inside $10\re$ (see equation 51 in CO07).  At
  late times the thermal X-ray luminosity is $\sim 10^{40}$ erg/s and
  the post-starburst (E+A) luminosity is a fraction of a percent of
  the light from stars.}
\label{fig:lxb}
\end{figure}

The temporal evolution of the integrated luminositites of other mass
components of the models clearly reflects the accretion history, as
shown in Figure~\ref{fig:lxb}. In panels $a$ we show the evolution of
the X-ray luminosity of the hot galaxy atmosphere, integrated within
$10\re$. The sharp peaks are due to sudden increases in the X-ray
surface brightness profiles in the central regions, consequence of AGN
feedback. During more quiescent phases, $\lx$ attains values
comparable to the observed ones, and it is expected that a central
galaxy will reach higher values, due to confining effects of the
ICM. On the contrary, stripping effects of the ICM in satellite
galaxies lead to a further reduction (see Shin et al.~2010b).  In
panels $b$ we show the reprocessed IR radiation from the central
starbursts and the associated SMBH accretion, with peak values similar
to those observed in ULIRGs (e.g., see Nardini et al.~2010).  We note
that peaks of nuclear IR emission coupled with nuclear radio/X-ray
emission have been recently reported in a sample of elliptical
galaxies (Tang et al.~2009); the authors suggest that the correlation
indicates that the excess IR emission is related to nuclear activity,
and presumably due to hot dust heated by the central AGN.  Finally, in
panels $c$ the luminosity of the starburst in the UV and optical,
after correction for absorption (i.e., as would be seen from infinity,
as described in CO07) is shown. It is apparent how a large fraction of
the starburst luminosity output occurs during phases when shrouding by
dust is significant (e.g., Rodighiero et al.~2007, Brusa et al.~2009),
i.e., the model would be observed as an IR source with UV and optical
in the range seen in brighter E+A sources. Opacity effects are also
apparent in the rise of the two luminosities in model \BIIIzd~at late
times, due to a substantial decrease in the ISM opacity associated
with the degassing of the galaxy central regions consequent to the
last major burst around 10 Gyr.  Of course, due to the longer
time-scales of star formation, the peaks of the UV and optical light
curves are less sharp than those relative to SMBH luminosity; we
notice that the measure of the different time-scales of nuclear
accretion and associated star formation can now be measured, with very
interesting results (Wild, Heckman \& Charlot 2010).

\begin{figure}
\includegraphics[angle=0,scale=0.8]{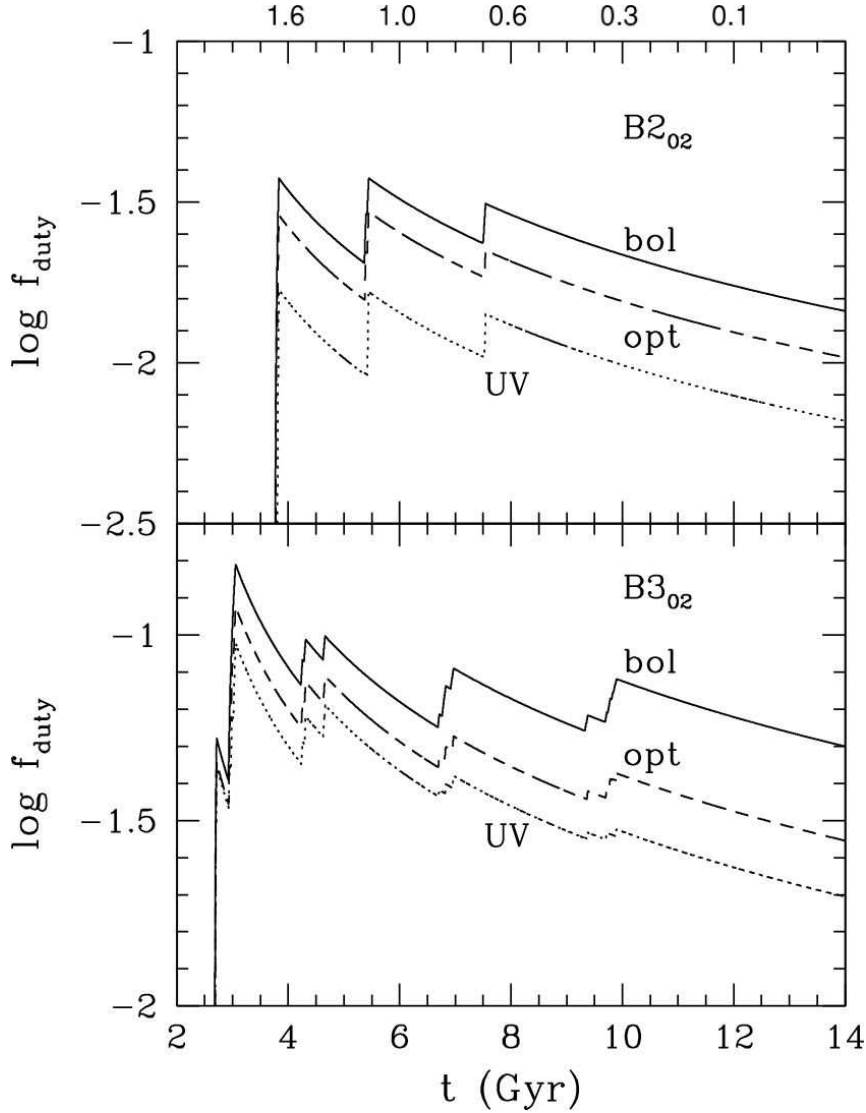}
\caption{Time evolution of cumulative duty-cycle (defined as fraction
  of the elapsed time during which the AGN is flaring at more than a
  fixed threshold) of the nucleus in the bolometric, in the absorbed
  (i.e., as would be seen from infinity) optical, and absorbed UV
  bands, for models \BIIzd~and \BIIIzd. For example, in the optical
  band between 1\% and 3\% of SMBH would be seen in the ``on'' state
  at the current epoch, if using as a time baseline for the estimate
  the whole integration time of the simulation (see the text for a
  more detailed explanation). The scale at the top indicates the
  redshift in a flat universe corresponding to the age in the
  abscissae axis.}
\label{fig:duty}
\end{figure}

As anticipated, at variance with our previous papers, we compute here
the duty-cycle as the total time spent by the AGN at high luminosity
phases, normalized to the age of the system at the specified time,
while in our previous papers we adopted a luminosity-weighted
definition. The new duty-cycle values (at the current epoch) are just
obtained by dividing the figures in Column 14 of Table 1 by the total
time spent by the simulation, i.e., 12.5 Gyrs.  In other words, we
estimate the observable duty-cycle as the fraction of the total time
that the AGN is in the ``on'' state. We found that the average values
are very similar to the luminosity-weighted values, but here we prefer
to use the new approach, as its interpretation is more direct (the
older one depending also on the adopted time-fraction of the total
time interval over which backward integration is done). The results in
different wavebands (namely, optical and UV after absorption, and
bolometric) are presented in Figure~\ref{fig:duty}: as expected, at
each time the larger duty-cycles are in the bolometric, followed by
(absorbed) optical and finally by absorbed UV.  For example, for the
two best models the total time spent at high luminosity in optical is
in the range $130-380$ Myr.  The bolometric values obtained from Table
1 are comparable (even though slightly larger) than the duty-cycles
computed according to the luminosity-weighted recipe adopted in CO07
and in Paper I. For example, we see that model \BIIIzd~has a
bolometric duty-cycle of $\simeq 5\%$, while in the absorbed UV the
duty-cycle drops to $\simeq 2\%$.

Of course, these values (by construction) cannot take into account the
temporal decline of the accretion activity over the Hubble time.  As
an experiment, we considered the duty-cycle obtained by starting the
analysis at 6 and at 9 Gyrs. The resulting numbers are significantly
smaller: for example, in model \BIIzd~the duty-cycle is zero when
starting at 9 Gyr (as no bursts occur after $\simeq 7.5$ Gyr), while
using 6 Gyrs as initial time we obtain $\simeq 5.2\,10^{-3}$,
$2.6\,10^{-3}$, and $2.5\,10^{-3}$ in bolometric, optical, and UV
bands, respectively. A similar (but less strong) reduction is also
presented by model \BIIIzd, with computed duty-cycles in the
bolometric, optical, and UV of $\simeq (4\,10^{-2}, 1.5\,10^{-2},
7.5\,10^{-3})$ respectively (when starting the count at 6 Gyr),
and of $\simeq (3.6\,10^{-2}, 1.3\,10^{-2}, and 5.2\,10^{-3})$ (when
starting the count at 9 Gyr). These values compare nicely with observational 
estimates (e.g., Heckman et al.~2004, Greene \& Ho 2007).

\begin{figure}
\includegraphics[angle=0,scale=0.8]{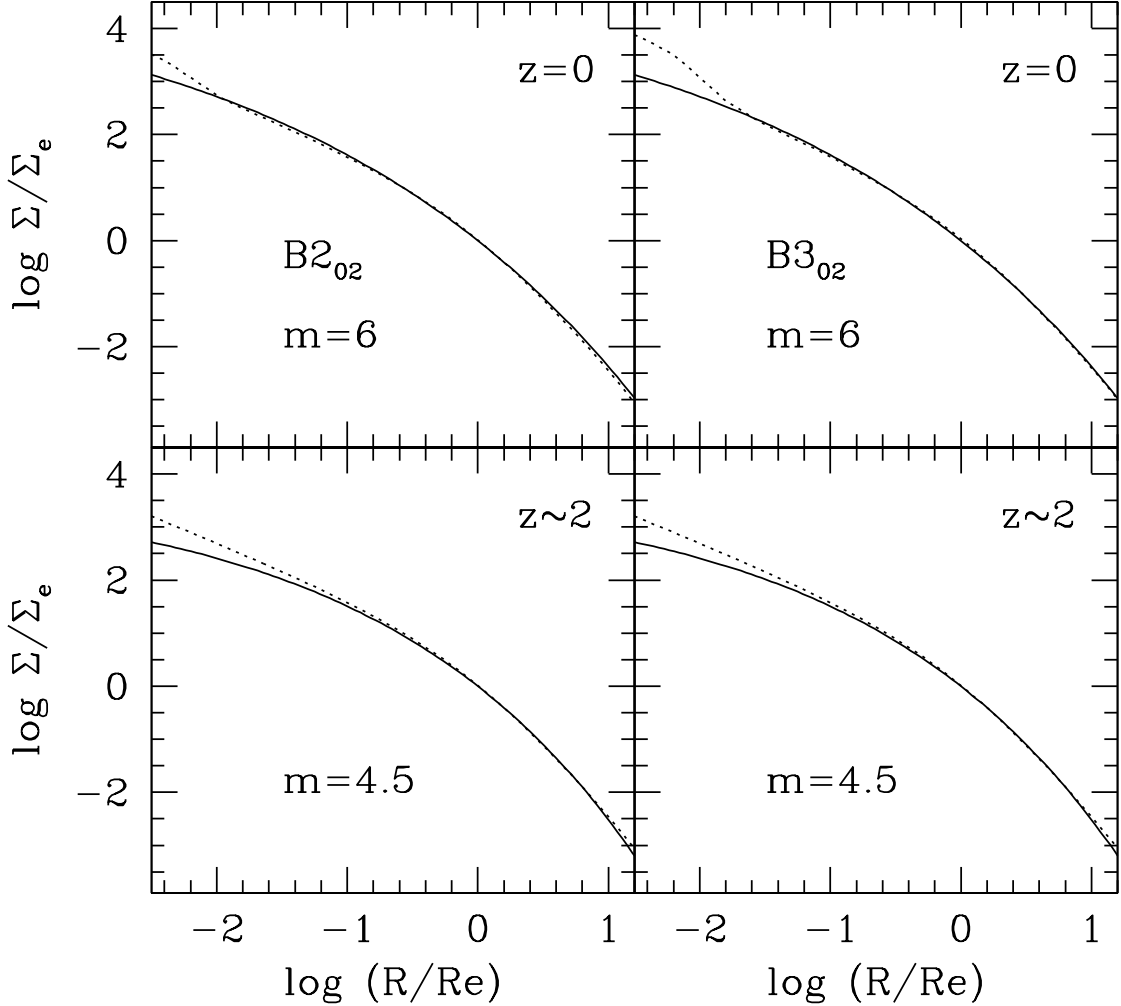}
\caption{Normalized projected stellar density profiles of models
  \BIIzd~(left panels, dotted lines) and \BIIIzd~(right panels, dotted
  lines), near the beginning (bottom panels) and at the end (top
  panels) of the simulations. Solid lines show the best-fit global
  Sersic profiles, and the corresponding best-fit index is given. In
  model \BIIzd~the effective radius remains almost unchanged at
  $\re=6.9$ kpc, and the projected mass density is
  $\Sigma(\re)\simeq 215\Msun$ pc$^{-2}$. In model \BIIIzd~the
  effective radius contracts from $\re=6.9$ kpc to $\re=6.2$ kpc,
  while $\Sigma(\re)$ increases from $\simeq 215\Msun$ pc$^{-2}$
  to $\simeq 263\Msun$ pc$^{-2}$.}
\label{fig:surf}
\end{figure}

We now focus on a remarkable observational features of our models,
already noticed in the case of purely radiative feedback models in
CO07, and nicely confirmed by combined models.  The violent star
formation episodes associated with the recurrent nuclear accretion
events (with SMBH accretion to star formation mass ratios $\sim
10^{-2}$ or less) are induced by accretion feedback, and are spatially
limited to the central $10-100$ pc; thus, the bulk of gas flowing to
the center is consumed in the starburst. These findings are nicely
supported by recent observations (e.g, see Lauer et al.~2005; Davies
et al.~2007, 2009; Shapiro et al.~2010). In fact, as noticed by Lauer
et al.~(2005), where colors and luminosities of the nuclear regions of
elliptical galaxies are studied, on average the ``nuclear'' clusters
are bluer than the surrounding galaxy, and in some case they are very
well-resolved at the distance of Virgo.  An interesting example is the
central system in NGC 4365 (Fig.~3 in Lauer et al.~2005), in which a
blue, extended source well interior to the core is
detected\footnote{It is interesting that small compact blue nuclear
  clusters are seen in many ellipticals with cores (T. Lauer, private
  communication)}.
Note that the ``age'' effect of the new stars
on the global stellar population of the galaxy is small, as the new
mass is only 3\% of the original stellar mass, and it is virtually all
accumulated during the first Gyrs, so that the mass-weighted age of
the final model is still of the order of 12 Gyrs. The half-mass radius
of the final stellar distribution (without considering adiabatic
contraction, nor the reduction of the stellar mass distribution due to
galactic winds, two phenomena not considered in the present
simulations) remains almost unchanged in model \BIIzd, in accordance
with the moderate amount of new stars formed (Figure~\ref{fig:fmb202},
red line, top panel; and Table 1), while it contracts by $\sim 10\%$
in model \BIIIzd~(due to the larger amount of gas transformed in new
stars, see Figure~\ref{fig:fmb302}, red line, top panel; Table 1),
from $\re=6.9$ kpc to $\re=6.2$ kpc.  This addition of the new stars
in the central regions of the galaxy is made apparent in
Figure~\ref{fig:surf}, where with dotted lines we show the initial
(bottom panels) and spatial (top panels) projected stellar density
profiles of models \BIIzd~and \BIIIzd, together their best-fit (solid
lines) obtained with the Sersic (1968) law
\beq 
\Sigma(R)=\Sigma_0{\rm e}^{-b(R/\re)^{1/m}},
\eeq 
(where $b = 2m-1/3+4/405m+{\cal O}(m^{-2})$, Ciotti \& Bertin 1999).
As expected, the profiles show an increase of the global best-fit
Sersic parameter $m$ from $\simeq 4.5$ up to $m\simeq 6$, due to the
mass accumulation in the central regions. Remarkably, the final value
of $m$ is within the range of values commonly observed in massive
ellipticals: however, in the final model we note the presence of a
central ($\sim 30$ pc) nucleus originated by star formation which
stays above the best fit profile, similar to the light spikes
characterizing ``nucleated'' or ``extra-light'' galaxies, that are
usually attributed to galaxy merging (e.g., see Hopkins et al.~2009,
and references therein).  Such ``cuspy'' ellipticals provide one of
the two important branches of the elliptical galaxy sequence (e.g.,
see Graham 2004, Graham \& Driver 2005, Lauer et al.~2005, Kormendy et
al.~2009; see also Ciotti 2009b).  Observational evidence is also
accumulating that the central parts are quite metal rich (e.g., see
Chilingarian, De Rijcke, \& Buyle 2009; Lee et al.~2010; Rafanelli et
al.~2010) as would be expected if the origin were from infalling gas
recycled from evolving stars.

\begin{figure}
\includegraphics[angle=0,scale=0.8]{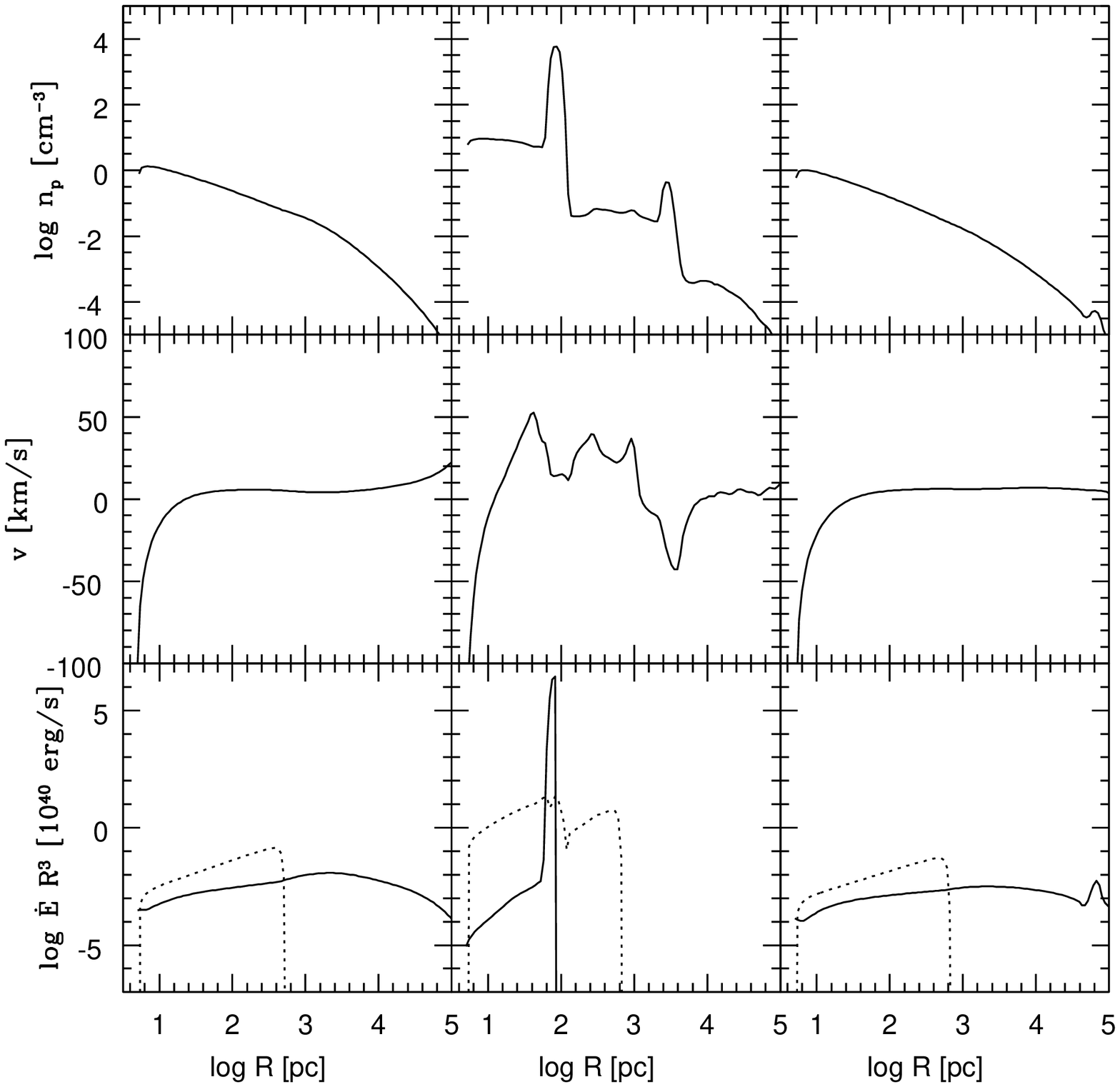}
\caption{Volume weighted feedback energy deposition just before (left
  row), at the end (central row), and after (right row) the first
  burst of model \BIIIzd~(bottom panels). Solid lines refer to
  radiative feedback and solid lines to mechanical feedback.  The
  radial profile of the ISM number density (top panels) and velocity
  (middle panels) are also shown.  Much of the radiation is absorbed
  in the cold shell at $\approx 100$ pc at the peak of the burst, but at
  other times the radiative input is broadly distributed. The
  mechanical energy input from the outflowing wind tends to be more
  centrally concentrated.}
\label{fig:dep}
\end{figure}

As in these models we are now considering both the effects of
radiative and mechanical feedback, it is interesting to show where the
bulk of feedback energy is deposited. We illustrate this in
Figure~\ref{fig:dep}, for three times just before (left column), near
the end (central column), and after the first burst (right column) of
model \BIIIzd. For illustration, we also plot the corresponding radial
profiles of gas density (top panels) and gas velocity (central
panels). In the bottom panels solid lines represent the
volume-weighted profiles of radiative energy deposition, while the
dotted lines show the corresponding mechanical energy deposition. A
clear trend is apparent: while after and before the burst the
radiative feedback affects all the galaxy volume (due to the low
opacity of the hot ISM), and the mechanical feedback is concentrated
in the central kpc region, during the burst almost all the AGN
radiation is absorbed (and reprocessed in the IR) by the cold and
optically thick collapsing shell. This means that during the burst,
mechanical feedback plays a major role in the model evolution, a
conclusion that seems to be supported also by observations (Moe et
al.~2009, Dunn et al.~2010. We also note that the values deduced for
the mechanical feedback efficiency during the burst nicely match the
range adopted in the present study, as detailed in Table 1).

We note the general properties of all of the models we have
studied. If there is proper allowance for mass conservation, then
constraining the models to allow for appropriate SMBH growth {\it
  requires} a mechanical efficiency at or below $3\,10^{-4}$, at least
a factor of ten lower than is commonly assumed.  Such a relatively low
mechanical efficiency can mean that even outflows driven from as far
as 0.1 pc from the SMBH can be important (e.g., Proga, Ostriker \&
Kurosawa 2008; Kurosawa \& Proga 2009; Kurosawa, Proga, \& Nagamine
2009).

\subsection{The ``best'' models with the explicit time-dependent term
  in the nuclear wind differential equation}

\begin{figure}
\includegraphics[angle=0,scale=0.8]{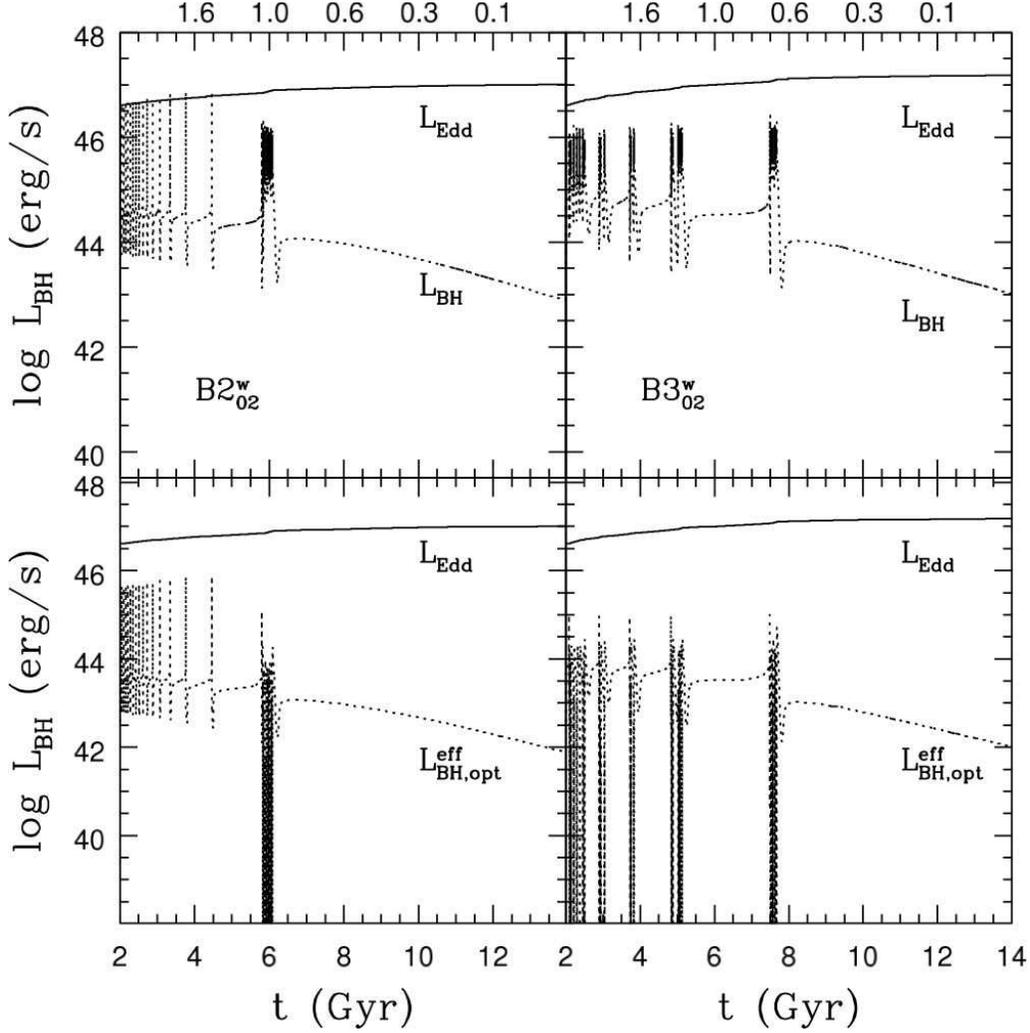}
\caption{Luminosity evolution of time-dependent wind models
  \BIIzdw~(left panels) and \BIIIzdw~(right panels).  Model quantities
  are given in Table 1, and lines are as in Figure~\ref{fig:flb}.
  Comparing these results to those shown in Figure~\ref{fig:flb} we
  see qualitatively similar behaviour but an even lower duty-cycle in
  the on-state, an even more dramatic decline in bursting behavior
  with decreasing $z$ and an even lower final luminosity.}
\label{fig:flbw}
\end{figure}

Before concluding, we briefly illustrate the effects of the inclusion
in the simulations of the explicit time-dependent term in the nuclear
wind differential equation (equation 29 in Paper I): we recall that
this term takes into account the finite propagation time of the
nuclear wind.  As can be seen by comparison of Figure \ref{fig:flbw}
with Figure \ref{fig:flb}, the evolution is somewhat different from
the analogous models in which the time-dependent term is
suppressed. In general, the temporal structure of the bursts is more
complex, a consequence of the additional time-propagation scale of the
mechanical feedback, which is out of phase with the global evolution
of the galactic gas flows. In fact, as can be seen from Column 13 in
Table 1, these B$^{\rm w}$ models produce a much larger number of
bursts than the corresponding B models. However, the total time spent
in the high-luminosity state is considerably shorter, as the
characteristic burst duration is now (for example) $\sim 1$ Myr for models
\BIIzdw~and \BIIIzdw, rather than $\sim 30$ Myr for the models
\BIIzd~and \BIIIzd. Overall, the cumulative duty-cycle for the new
models is reduced by roughly a factor of 10, with $f_{\rm duty}\approx
0.004$.

From the point of view of the central SMBH, the primary difference is
the greater effectiveness of feedback in shutting off accretion, so
that the growth of the SMBH is now reduced with respect to B models,
and more gas is retained in the galaxy producing an X-ray coronal
luminosity 10 times larger.  From the point of view of the final
stellar distribution, the present models behave similarly to the
models described in the previous Section, i.e., at the end of the
simulation a central ``cusp'' of new stars is produced, in qualitative
agreement with observations. We finally notice how the
luminosity-weighted BLR wind solid opening angle (Table 1, Column 12)
for this last set of models is found in good agreement with the most
recent observational estimates (Lu et al. 2010).

\section{Discussion and conclusions}

In this paper we have addressed, with the aid of 1D hydrodynamical
simulations, the combined effects of radiative and mechanical feedback
from the central SMBH on the gas flows in elliptical galaxies.  The
investigation is in the line of previous papers, where the input
physics and the galaxy models have been substantially improved over
time.  In Papers I and II we focused on purely radiative and purely
mechanical feedback models, and for both the cases we found
difficulties in reproducing some of the most important observational
features of observed galaxies.  In the present paper we explored the
behavior of {\it combined} models, i.e. when both a physically
motivated implementation of radiative and mechanical feedback effects
is active in the code.

We briefly recall here the main secure points on which our framework
is based. First of all, it is known from stellar evolution theory, and
firmly supported by observations, that the recycled gas from dying
stars is an important source of fuel for the central SMBH, both in its
amount (summing up to 20-30\% of the total mass in stars) and in its
availability over cosmic epochs.  It is also obvious that the recycled
gas, arising from stars in the inner several kpc of the galaxy
(assumed a giant elliptical), is necessarily a subject of a classical
radiative cooling instability, leading to a collapse towards the
center of metal rich gas. As a consequence, star-bursts occur and also
the central SMBH is fed. The details of how much is accreted on the
central SMBH vs. consumed in stars vs. ejected from the center by
energy input from the starburst and AGN are uncertain.  But order of
magnitude estimates would have the bulk going into stars or blown out
as a galactic wind, with a small amount going into the central SMBH.
In addition, since at the end of a major outburst, an hot bubble
remains at the center, both processes shut themselves off, and it will
take a cooling time for the cycle to repeat. In other words,
relaxation oscillations are to be expected, but their detailed
character is uncertain.  Finally, order of magnitude estimates would
indicate that during the bursting phase the center would be optically
thick to dust, so one would observe a largely obscured starburst and
largely obscured AGN with most radiation in the far IR; as gas and
dust are consumed, the central sources become visible. Much of the AGN
output occurs during obscured phases: then there is a brief interval
when one sees a ``normal'' quasar, and finally one would see a low X-ray
luminosity and E+A spectrum galaxy, with A dominating in the central
several hundred pc for $10^{7-8}$ yrs. Such figures are consistent
with observed statistical occurrence of E+A galaxies in homogeneous
samples that are nowadays available (e.g., Goto 2007).

In sum, we find that inclusion of a modest amount of (momentum and
energy driven mechanical) feedback significantly improves the
correspondance with reality, with the optimal level being
$<\epsw>\simeq 10^{-4.5}$. In the models where we took feedback to be
$10^2$ more efficient, as is commonly assumed, and assuming
conservation of both mass and momentum, we found that the results were
unsatisfactory with too low a SMBH mass growth and too low thermal
X-ray luminosity from the gas. The logic is simple. Very efficient
mechanical feedback keeps gas from infalling to the central SMBH and
strips the galaxy of thermal gas.  In order to better study the
problem, we presented three different versions of the same models,
namely combined models (type A) in which the mechanical energy
associated with the nuclear wind is computed under the assumption of a
fixed opening angle for the wind cone; combined models (type B) in
which the opening angle of the nuclear wind and the mechanical
feedback efficiency are function of the Eddington normalized
istantaneous bolometric luminosity of the SMBH. Finally, in the third
family we briefly consider B models with the full differential
equation of mechanical feedback, in which also the propagation
velocity of the nuclea wind is considered.  In each family of models
we assumed different (but physically plausible) combinations of
parameters.

Overall, we have confirmed the results in CO07, and completed the
model investigation in Papers I, II and V. More in detail, the main
results can be summarized as follows:

1) Radiative heating (primarily due to X-rays) without any mechanical
energy input greatly reduces the ``cooling flow catastrophe'' problem,
but leads to a result that is still defective as compared to detailed
observations of local elliptical galaxies in that the central SMBH
would be too bright and too massive and the galaxy would be too blue
at $z=0$ (CO07, Paper I).

2) Utilizing mechanical energy alone from an AGN wind with fixed
efficiency can address the problems but does not give a solution that
in detail satisfies the observations (Papers I and II).  If the chosen
efficiency is large, then we obtain (consistent with Di Matteo et
al.~2005 and others) a giant burst, and a not too large SMBH but do
not get any late time AGN and the overall duty cycle is too small. If
the fixed efficiency is made low enough to avoid these problems, then
one simply reverts to case (1) above, the radiative case. Purely
mechanical models with luminosity-dependent mechanical efficiency
(which seems to be what is indicated both by observations and detailed
multi-dimensional hydrodynamical simulations of radiatively driven
winds by Proga 2007, Proga et al.~2008, Kurosawa \& Proga 2009,
Kurosawa et al.~2009) perform better.

3) The combined models explored in this paper, in which both radiative
and mechanical feedback are allowed are clearly better than the two
limit cases described above. This family of models, with mechanical
energy efficiency proportional to the luminosity, when combined with a
realistic treatment of the radiative effects, does seems it to be
consistent with all observations for a range of efficiencies $\epsw$
which includes the values thought to obtain either from an analysis of
the observations or from theoretical modeling of the central
engines. And, of course, actual observations indicate that both
feedback processes do occur: both the radiative output and the
broad-line and narrow-line winds are observed from AGNs (e.g.,
Alexander et al.~2009, 2010).

4) It is found that radiative and mechanical feedback affect different
regions of the galaxy at different evolutionary stages. During the
``quiescent'' phases, when the ISM is optically thin, radiative
heating is distributed over all the galaxy body, while the mechanical
feedback is deposited in a region of a kpc scale radius. This produces
a characteristic feature which is not present in purely radiative
models, i.e. the density profile is flatter within $\sim 1$ kpc than in
models without mechanical feedback. This in turn produce a more flat
core in the observed X-ray surface brightness profile (Pellegrini,
Ciotti \& Ostriker 2010, Paper IV, in preparation).  However, during
the burst, the collapsing cold shells are optically thick, and most of
the radiation is intercepted and re-radiated in the IR. It is found
that during these short phases actually is the mechanical feedback
that is playng the major role. This means that a proper description of
SMBH feedback requires both the physical components, as discussed in detail 
in Ostriker et al.~2010.

\begin{figure}
\hskip -3truecm
\includegraphics[angle=-90,scale=0.75]{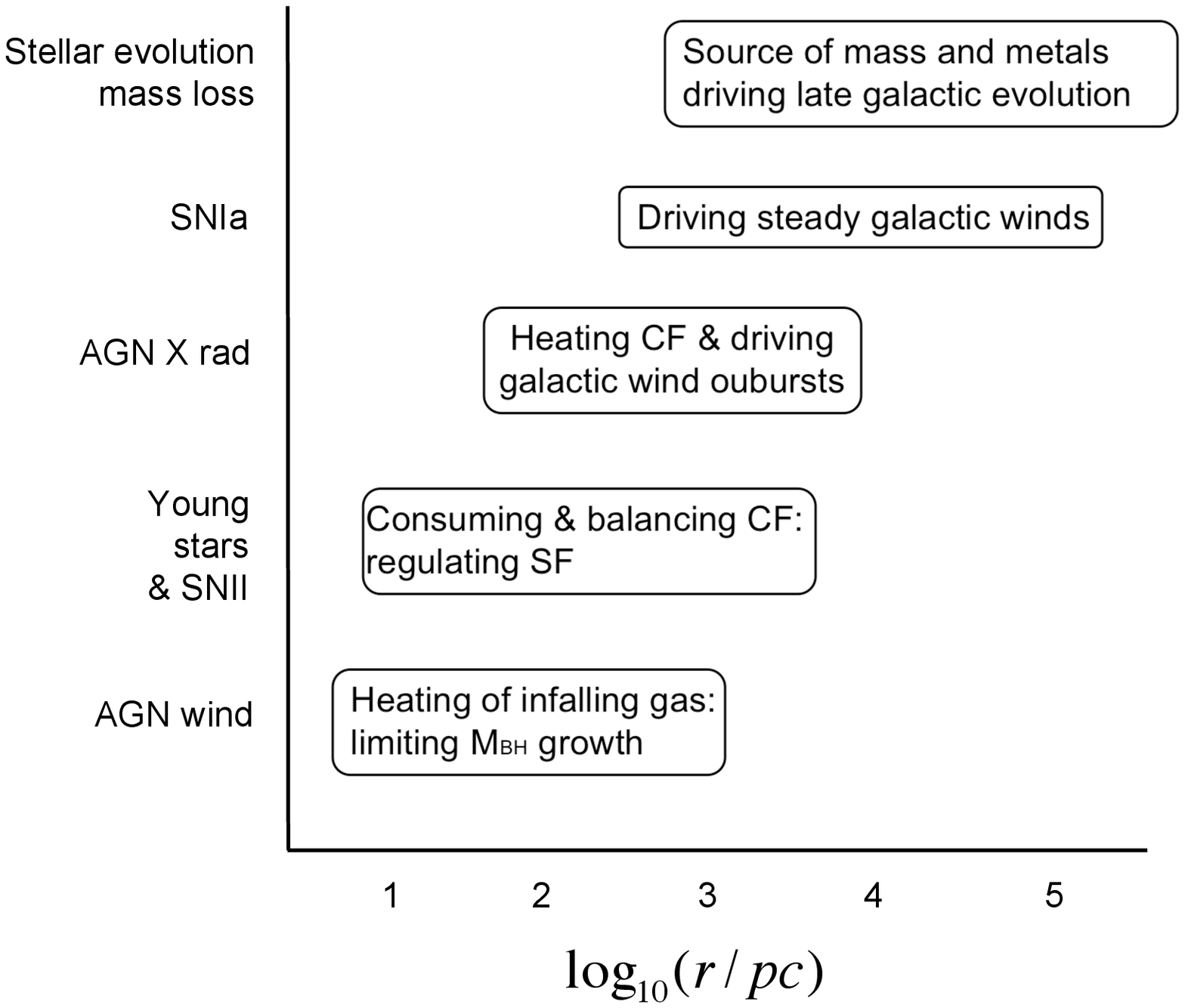}
\caption{This diagram shows the radial intervals where the different
  physical mechanisms affect the evolution of gas flows driven by
  stellar evolution, in a representative (isolated) elliptical galaxy.}
\label{fig:scheme}
\end{figure}

We summarize the overall situation with a cartoon in Figure
\ref{fig:scheme} in which we indicate the spatial regions within which
each physical mechanism is dominant. Since the detailed hydro is
highly time-dependent, the cartoon greatly oversimplifies the complex
situation but shows the regions where a process is most important
during the time interval that it is important.

Tests of the overall picture presented in these papers are numerous
and obvious.

i) The E+A spectrum should be found to come from $\lsim 100$ pc
regions, be younger and be significantly more metal rich than the bulk
of the stars in a given galaxy.

ii) The duration of the AGN bursts should be quite short with regard
to cosmic time, in the range of burst duration say $0.3\lsim \Delta
t\lsim 30$ Myrs.

iii) Since satellite galaxies in clusters have more difficulty in
retaining the recycled gas than do central galaxies and isolated
galaxies, the E+A phenomenon should be rarer in these systems, the
central stellar cusps should be weaker and the incidence of the AGN
phenomenon rarer.

iv) The fraction of the time that normal elliptical galaxies spend in
the bursting state should be small and fairly steeply declining with
increasing cosmic time. The final SMBH luminosities will be typically
$\sim 10^{-5}\ledd$.

v) If cooling instabilities in recycled gas dominate (at late times)
the fueling of AGN bursts, then evidence for merging activity should be
relatively rare and the gas seen during the outbursts should be
relatively rich in metals (including S-process elements).

We conclude by recalling that we restricted our study for simplicity
to the case of an isolated elliptical galaxy, therefore excluding the
confining or stripping effect of the ICM on the galactic X-ray
emitting corona. The two physical phenomena have opposite effects.  On
one hand, as shown in Shin et al. (2010b), the ram-pressure stripping
of the ICM on the galactic atmosphere has the effect of reducing the
global X-ray luminosity of the galaxy, and so also to retard or even
suppress the cooling catastrophe and the associated AGN bursts. In
turn, this will also reduce the starburst activity that we
(invariably) find associated with strong bursts. On the other hand,
galaxies residing in the central regions of a cluster are presumably
more affected by external pressure effects than by ram pressure
stripping. Such galaxies would be on average more X-ray luminous, and
show not only larger duty-cycle values, but also younger and bluer
nulear cusps near the SMBH, on the 100 pc scale, than similar galaxies
orbiting in gthe cluster outskirts.  Therefore, we expect that the
coronal X-ray luminosity of the hot gas, and the number of central
bursts of the models presented in this paper, are lower limits for the
case of galaxies immersed in a realistic ICM in the central regions of
a cluster. Also, the effect of mechanical feedback due to a jet, which
is relevant at low accretion luminosities (i.e., during the hot phase
accretion common at late times, e.g. see Allen et al.~2006, Merloni \&
Heinz 2007), is not included. The associated reduction of the
accretion luminosity in the low-luminosity states will bring our
models nearer to the observed Eddington ratios of low-luminosity AGNs.
However, we stress that the maintenance of the SMBH masses to the
observed level, in presence of the important amounts of recycled
metal-rich gas produced over an Hubble time by the aging stellar
populations, is mainly due to the combined effect of the feedback
terms included in the simulation, i.e. SNIa and SNII heating,
radiative heating, and nuclear wind feedback.

The (metal rich) recycled gas from stellar evolution is present {\it
  even in absence of external phenomena such as galaxy merging or
  input from cold gas flows}, that are often considered as the natural
way to induce QSO activity.  Therefore, one of the main results of our
simulations (also considering all the simplifications in the treatment
of physics, and of the geometry of the code), is that {\it the
  evolution of an isolated galaxy, subject to internal evolution only,
  can be quite complicated} (e.g., see Pierce et al.~2007), and that
AGN feedback will lead to central AGN and starburst activity in many
widely spaced brief intervals (e.g., Shi et al.~2009). We note that
the possibility of QSO activity even in absence of merging, has also
been recently proposed also by others (e.g., Li et al.~2008, Kauffmann
\& Heckman 2009, Tal et al.~2009).

Clearly, the main limitation of the models explored in our series of
papers is the adopted spherical symmetry. This choice has been
necessary as the main focus of the study has been the understanding of
the physics behind AGN feedback. The necessity to explore the
parameter space, and the time-expensive numerical integration of
heating/cooling, star formation, and radiative trasport equations from
the pc scale near the SMBH up to the hundred-kpc scale of the galaxy
outer regions, in presence of multiple mutually interacting shocks,
forced us to use a 1-D code. However, we are now working on 2-D
simulations, that allows for a more realistic description of
mechanical feedback, of the cold shell stability and fragmentation,
and of the effect of angular momentum of the accreting gas.  Work
performed to date indicates that several aspects of the gas evolution
are quite similar in the 2-D and 1-D simulations.

\acknowledgments We thank an anonymus Referee for useful comments and
suggestions that improved the paper. We also thank Sudeep Das for
preparing Figure \ref{fig:cartoon}, and Jenny Greene, Tim Heckman,
Guinevere Kauffmann, Silvia Pellegrini and Eliot Quataert for
interesting discussions. D.P. acknowledges support by the National
Aeronautics and Space Administration under Grant/Cooperative Agreement
No~NNX08AE57A issued by the Nevada NASA EPSCoR program.

\clearpage
\begin{deluxetable}{lcccccccccccccc}
\rotate
\tablecaption{Properties of computed models}
\tabletypesize{\scriptsize}
\tablewidth{0pt}
\tablehead{
\colhead{Model}&
\colhead{$\epswM$}&
\colhead{$<\epsw>$}&
\colhead{\tablenotemark{a}$<\epsj>$}&
\colhead{$<\eps_{\rm EM}>$}&
\colhead{$\log \Delta\mbh$}&
\colhead{$\log \Delta M_*$} &
\colhead{$\log \Delta M_{\rm w}$}&
\colhead{$\log \mgas$} &
\colhead{$\log l^{\rm eff}_{\rm BH,opt}$} &
\colhead{$\log L_{\rm X,ISM} $} &
\colhead{$<\Delta\Omega_{\rm w}>$}&
\colhead{$N_{\rm b}$} &
\colhead{$\Delta t_{\rm b}$} \\
\colhead{(1)}&
\colhead{(2)}&
\colhead{(3)}&
\colhead{(4)}& 
\colhead{(5)}&
\colhead{(6)}&
\colhead{(7)}&
\colhead{(8)}&
\colhead{(9)}&
\colhead{(10)}&
\colhead{(11)}&
\colhead{(12)}&
\colhead{(13)}&
\colhead{(14)}&
}
\startdata
\hline

{\bf A0} &$5\,10^{-3}$   &$5\,10^{-3}$     &$1.2\,10^{-2}$    &0.003  &7.17  &6.43  & 10.38 & 7.65 & -7.71 & 36.64& --   & 0&   0\\
{\bf A1} &$2.5\,10^{-4}$ &$2.5\,10^{-4}$   &$4.4\,10^{-3}$    &0.053  &7.74  &9.70  & 10.34 & 9.05 & -7.59 & 39.13& 0.5  & 9&32.2\\
{\bf A2} &$10^{-4}$      &$10^{-4}$        &$2.95\,10^{-3}$   &0.062  &7.97  &9.80  & 10.28 & 9.67 & -7.59 & 40.16& 0.5  &18&65.0\\
{\bf A3} &$5\,10^{-5}$   &$5\,10^{-5}$     &$1.04\,10^{-3}$   &0.078  &8.48  &10.21 & 10.34 & 9.44 & -7.79 & 39.73& 0.5  &23&269.5\\
\hline
\hline
{\bf B0} &$5\,10^{-3}$   &$6.3\,10^{-5}$   &$3.3\,10^{-3}$    &0.043  &8.82  &9.76  & 10.26 & 9.72 & -5.67  & 40.37&0.066& 5&  90.6\\
{\bf B1} &$2.5\,10^{-3}$ &$4.3\,10^{-5}$   &$3.2\,10^{-3}$    &0.047  &8.87  &9.78  & 10.29 & 9.64 & -5.77  & 40.16&0.069& 6& 122.5\\
{\bf B2} &$10^{-3}$      &$2.3\,10^{-5}$   &$2.9\,10^{-3}$    &0.052  &8.96  &9.84  & 10.34 & 9.44 & -5.96  & 39.78&0.077&12& 205.0\\
{\bf B3} &$3\,10^{-4}$   & $1.3\,10^{-5}$  &$2.1\,10^{-3}$    &0.062  &9.25  &10.20 & 10.38 & 9.32 & -6.05  & 39.62&0.108&27& 391.1\\
\hline
{\bf \BZzd}   &$5\,10^{-3}$   &$3.9\,10^{-5}$ &$2.1\,10^{-3}$ &0.090 & 8.62 & 9.49 & 10.28 & 9.68 & -5.21  & 40.21&0.066&1& 36.9\\
{\bf \BIzd}   &$2.5\,10^{-3}$ &$4.2\,10^{-5}$ &$1.6\,10^{-3}$ &0.105 & 8.75 & 9.83 & 10.32 & 9.27 & -5.44  & 39.54&0.074&6& 132.1\\
{\bf \BIIzd}  &$10^{-3}$      &$2.0\,10^{-5}$ &$1.9\,10^{-3}$ &0.105 & 8.77 & 9.74 & 10.27 & 9.68 & -5.13 & 40.23&0.093&5&  172.8\\
{\bf \BIIIzd} &$3\,10^{-4}$   &$1.2\,10^{-5}$ &$1.2\,10^{-3}$ &0.133 & 9.06 & 10.22 & 10.31 & 9.34 & -5.43 &39.63&0.080&22& 600.0\\
\hline
{\bf \BZw}    &$5\,10^{-3}$   &$4.8\,10^{-5}$ &$5.0\,10^{-3}$ &0.028 & 8.55 & 9.41  &10.31 &9.50 & -5.79  & 39.94&0.172&93&  16.1\\
{\bf \BIw}    &$2.5\,10^{-3}$ &$2.4\,10^{-5}$ &$4.7\,10^{-3}$ &0.029 & 8.64 & 9.40  &10.25 &9.74 & -5.41  & 40.60&0.150&64&  18.2\\
{\bf \BIIw}   &$10^{-3}$      &$2.0\,10^{-5}$ &$2.5\,10^{-3}$ &0.047 & 8.91 & 10.21 &10.35 &9.36 & -5.97  & 39.68&0.176&192& 67.5\\
{\bf \BIIIw}  &$3\,10^{-4}$   &$6.5\,10^{-6}$ &$2.6\,10^{-3}$ &0.047 & 9.04 & 10.18 &10.32 &9.66 & -5.79  & 40.15&0.241&120& 95.0\\
\hline
{\bf \BZzdw}  &$5\,10^{-3}$   &$6.9\,10^{-5}$ &$3.3\,10^{-3}$ &0.068 & 8.35 & 8.97  &10.33 &9.37 & -5.25  & 39.65&0.208&46& 13.7\\
{\bf \BIzdw}  &$2.5\,10^{-3}$ &$4.0\,10^{-5}$ &$2.8\,10^{-3}$ &0.074 & 8.44 & 9.04  &10.32 &9.46 & -5.21  & 39.84&0.242&31& 12.9\\
{\bf \BIIzdw} &$10^{-3}$      &$2.9\,10^{-5}$ &$2.1\,10^{-3}$ &0.093 & 8.64 & 9.87 &10.32 &9.52 & -5.22  & 39.91&0.266&73& 67.1\\
{\bf \BIIIzdw}&$3\,10^{-4}$   &$4.9\,10^{-6}$ &$1.7\,10^{-3}$ &0.111 & 8.89 & 10.36 &10.33 &9.57 & -5.26  & 40.03&0.077&102& 423.6\\
\hline
\enddata

\tablecomments{Masses are in units of solar masses and luminosities in
  erg s$^{-1}$. In Type A models the nuclear wind efficiency is
  maintained constant, i.e., $\epsw=\epswM$, while in Type B models
  the value $\epswM$ is reached when $\lbh\geq 2\ledd$ (equation
  [6]). In models with the subscript $02$ the maximum radiative
  efficiency is $\epsz=0.2$ instead of 0.1 (equation [1]).  Mass
  accretion weighted efficiencies in Columns (3)-(5) are calculated
  according to equation (34) in Paper I. $\Delta\mast$ is the total
  amount of star formed during the model evolution, $\Delta M_{\rm w}$
  is the total amount of ISM lost at 10$\re$ and $\mgas$ the
  instantaneous amount of gas inside 10$\re$. The scaled luminosity
  $l^{\rm eff}_{\rm BH,opt}=\lbhefopt/\ledd$ in Column (10) is the
  Eddington normalized SMBH luminosity in the optical as would be seen
  at infinity after absorption, with $\lbhefopt=0.1\lbh$ at the first
  grid point (see CO07 for details).  Column (12) lists the
  luminosity-weighted solid opening angle of each of the two BLR wind
  conical regions (normalized to half solid angle). 
  In model A0 no value is provided, as the nuclear luminosity never exceedes 
  the limit $\ledd/10$.
  Finally, in
  Columns (13) and (14) we give the total number of bursts and the
  total time (in Myr) spent by the SMBH at (bolometric)
  $\lbh\geq\ledd/30$.}
\tablenotetext{a}{Figures corresponding to quantities calculated but
  not added to the hydrodynamical equations.}
\end{deluxetable} 

\clearpage

\end{document}